\documentclass[universe,article,accept,moreauthors,pdftex,10pt,a4paper]{mdpi}
\usepackage{amsmath}
\usepackage{hyperref}

\firstpage{1} 
\articlenumber{x}
\doinum{10.3390/------}
\pubvolume{xx}
\pubyear{2020}
\copyrightyear{2020}
\history{Received: 4 May 2020; Accepted: 5 June 2020; Published: 10 June 2020}

\Title{Was GW170817 a Canonical Neutron Star Merger? \\
	Bayesian Analysis with a Third 
	Family of Compact Stars}

\Author{{David Blaschke}$^{1,2,3}$\orcidC{}*, 
	Alexander Ayriyan$^{4,5}$\orcidA{}, David Edwin Alvarez-Castillo$^{1,6}$\orcidB{}, \\ Hovik Grigorian$^{4,5,7}$\orcidD{}}

\AuthorNames{David Blaschke, 
	Alexander Ayriyan, David Edwin Alvarez-Castillo, Hovik Grigorian}

\address{%
	$^{1}$ \quad Bogoliubov Laboratory for Theoretical Physics,
	Joint Institute for Nuclear Research,
	Joliot-Curie street 6,
	141980 Dubna, Russia; alvarez@theor.jinr.ru\\
	$^{2}$ \quad Institute of Theoretical Physics, 
	University of Wroclaw, 
	Max Born place 9, 
	50-204 Wroclaw, Poland\\
	$^{3}$ \quad National Research Nuclear University (MEPhI),
	Kashirskoe Shosse 31,
	115409 Moscow, Russia\\
	$^{4}$ \quad Laboratory of Information Technologies,
	Joint Institute for Nuclear Research,
	Joliot-Curie street 6,
	141980 Dubna, Russia; ayriyan@jinr.ru (A.A.); hovikgrigorian@gmail.com (H.G.)\\
	$^{5}$ \quad Computational Physics and IT Division, 
	A.I. Alikhanyan National Science Laboratory, 
	Alikhanyan Brothers street 2, 
	0036 Yerevan, Armenia\\	
	$^{6}$ \quad Institute of Nuclear Physics,
	Polish Academy of Sciences,
	Radzikowski street 152,
	31-342 Cracow, Poland \\
	$^{7}$ \quad Department of Physics, 
	Yerevan State University, 
	Alek Manukyan street 1, 
	0025 Yerevan, Armenia}

\corres{Correspondence: david.blaschke@uwr.edu.pl}

\abstract{
	We investigate the possibility 
	that GW170817 was not the merger of two conventional neutron stars (NS), but involved at least one if not two hybrid stars with a quark matter core that might even belong to a third family of compact stars. 
	To this end, we develop a Bayesian analysis method for selecting the most probable equation of state (EoS) under a set of constraints from compact star physics, which now also include the tidal deformability from GW170817 and the first result for the mass and radius determination for PSR J0030+0451 by the NICER
	Collaboration. We apply this method for the first time to a two-parameter family of hybrid EoS based on the DD2
	model with nucleonic excluded volume for hadronic matter and the color superconducting generalized nlNJL
	model for quark matter. The model has a variable onset density for deconfinement and can mimic the effects of pasta phases with the possibility of producing a third family of hybrid stars in the mass-radius diagram. The main findings of this study are that: 
	(1) the presence of multiple configurations for a given mass (twins or even triples) corresponds to a set of disconnected lines in the $\Lambda_1-\Lambda_2$ diagram of tidal deformabilities for binary mergers, so that merger events from the same mass range may result in a probability landscape with different peak positions; (2) the Bayesian analysis with the above observational constraints favors an early onset of the deconfinement transition, at masses of $M_{\rm onset}\le 0.8~M_\odot$ with an $M-R$ relationship that in the range of observed neutron star masses is almost indistinguishable from that of a soft hadronic Akmal, Pandharipande, and Ravenhall (APR) EoS ; (3) a few, yet fictitious measurements of the NICER experiment two times more accurate than the present value and a different mass and radius that would change the posterior likelihood so that hybrid EoS with a phase transition onset in the range $M_{\rm onset} = 1.1 - 1.6~M_\odot $ would be favored.
}

\keyword{Bayesian analysis; quark-hadron phase transition; pasta phases; hybrid compact stars; mass-radius relation; tidal deformability; GW170817; PSR J0030+0451; PSR J0740+6620
}

\usepackage{lipsum}
\usepackage{graphicx,pstricks,epsfig,rotating,amsmath,amssymb}

\newcommand{\bea}{\begin{eqnarray}}
\newcommand{\eea}{\end{eqnarray}}
\newcommand{\apgt} {\ {\raise-.5ex\hbox{$\buildrel>\over\sim$}}\ }

\begin{document}
	
	\section{Introduction}
	
	The detection of gravitational waves from the merger of two compact stars by the state-of-the-art interferometers on Earth has opened new possibilities of studying the equation of state (EoS) of matter under extreme conditions \cite{TheLIGOScientific:2017qsa}. 
	The process of the fusion of compact stars releases gravitational wave signals, which for the event GW170817 could be detected from the inspiralling phase when they probe the tidal deformability of the merging stars and allow deriving constraints on their radii \cite{Abbott:2018exr}.
	The gravitational waves from the postmerger phase could not be detected for this event. 
	However, once they become accessible to observation, the typical peak frequency in the Fourier spectrum of their ringdown signal will reveal information about the compactness of the hypermassive star that forms as a result of the fusion. 
	It has been suggested that both signals together would allow identifying a first-order phase transition in the neutron star merger 
	\cite{Bauswein:2018bma,Weih:2019xvw}.
	If deconfined quark matter can occur in neutron stars (NS) cores, then the fascinating question to be settled with NS observations is that for the onset mass $M_{\rm onset}$ of the deconfinement phase transition and whether the quark-core hybrid stars could form a separate family of compact stars. 
	It has been known since half a century that a third family of compact stars in the $M-R$ diagram would serve as an indicator of a strong phase transition in neutron-star matter \cite{Gerlach:1968zz}.
	
	In this work, we consider a class of hybrid star EoS based on state-of-the-art microphysical approaches to hadronic and quark matter phases, which are joined by an interpolating mixed phase construction that mimics the effects of pasta phases in the transition. 
	This theoretical approach is capable of describing hybrid stars as a third family of compact stars with varying $M_{\rm onset}$ and with the possibility to remove the gap in the $M-R$ diagram between this third family and the second family of canonical NS. 
	In order to address the above-mentioned goal of NS merger research, we compute the mass-radius and tidal deformability relations for our class of hybrid EoS 
	and perform a Bayesian analysis that takes into consideration the recent constraints from astrophysical observations. 
	
	The main important constraint for the EoS is the lower limit on the maximum mass of a compact star, which comes from the precise measurement of the mass for the most massive pulsars, like PSR J0740+6220 \cite{Cromartie:2019kug} and PSR J0348+0432 \cite{Antoniadis:2013pzd}.
	Fortunately, up to now, there are published data from two binary neutron star merger events denoted as GW170817 \cite{Abbott:2018exr} and GW190425
	\cite{Abbott:2020uma}.
	Since for the latter, no measurement of the tidal deformability has been reported, we shall not include it in our study. 
	Furthermore, the recent X-ray timing observation of the isolated millisecond pulsar PSR J0030+0451 performed by the NICER 
	(Neutron star Interior Composition ExploreR) experiment has provided a simultaneous measurement of both the mass and radius for that object \cite{Miller:2019cac,Riley:2019yda}. 
	
	Before describing our approach and its results in detail, we want to give a short introduction that relates our work to the actual status of research in this field.
	
	\subsection{EoS Constraints from $M-R$ Measurements before GW170817}
	Bayesian methods have proven to be a powerful tool for parameter estimation of a selected model using empirical data, as well as for statistical inference. 
	In the context of the determination of the neutron star equation of state, seminal works before the multi-messenger era include the analysis of X-ray bursters~\cite{Steiner:2010fz,Steiner:2012xt}, which unfortunately had to deal with uncertainties related to the atmospheric composition of the associated 
	neutron star. 
	Recent progress has been made using Bayesian methods for the mass and radius determination data of the Rossi X-ray Timing Explorer for some accretion-powered millisecond X-ray pulsars like, e.g., 4U 1702-429 \cite{Nattila:2015jra,Nattila:2017wtj}, SAX J1808.4-3658 \cite{Salmi:2018gsn}, and others.
	
	On the other hand, other Bayesian studies include the ones of Raithel et al.~\cite{Raithel:2017ity}, which introduced a popular piecewise multipolytrope EoS parametrization together with mass and radius measurements of several objects, whereas~\cite{Alvarez-Castillo:2016oln} included realistic
	EoS based on the hadronic density-dependent relativistic mean field (DD2) and the Nambu--Jona-Lasinio (NJL) quark matter model approaches to estimate the model parameters not only on already performed
	mass measurements, but also considering fictitious radius measurements on high-mass pulsars, which could provide evidence for mass twin stars. 
	In addition, Lackey and Wade~\cite{Lackey:2014fwa} performed a Bayesian inference study based on future detections of gravitation radiation from populations of inspiralling neutron stars using a piecewise polytrope EoS model with four polytropes 
	to conclude that the EOS determination above nuclear density could be significantly improved.

	\subsection{New Constraints on $M-R$ Relations in the Era of Multi-Messenger Astronomy}
	
	Following both the detection of gravitational radiation from the GW170817 event \cite{TheLIGOScientific:2017qsa} and the mass and radius measurement
	of the millisecond pulsar PSR J0030+0451 by NICER \cite{Riley:2019yda,Miller:2019cac}, a few works employing the Bayesian techniques appeared. 
	Since these new measurements are more precise and thus more reliable than previous ones, they allow for tighter constraints on the compact star EoS.
	Consequently, we base our present work on a study carried out in~\cite{Ayriyan:2018blj} where a Bayesian analysis considered
	EoS models featuring the effects of geometrical structures (pasta phases) at the hadron-quark interface on mass twins by
	using as inputs the tidal deformability estimates from GW170817 besides mass-radius measurements.
	This Bayesian analysis has recently been extended by including the NICER mass and radius measurement for PSR J0030+0451 
	\cite{Riley:2019yda,Miller:2019cac} in a study that used a purely hadronic EoS \cite{Alvarez-Castillo:2020aku,Alvarez-Castillo:2020fyn}.
	Regarding mass twins, a few other works have also recently studied their properties by varying model parameters with the result of finding disconnected branch structures in their deformability plots for compact star mergers~\cite{Christian:2018jyd,Montana:2018bkb}.
	Christian and Schaffner-Bielich~\cite{Christian:2019qer} used a hadronic relativistic mean-field EoS together
	with the simple constant speed of sound (CSS) model for the quark matter EoS to rule out strong phase transitions below $1.7n_0$ within
	their study. This strong claim should, however, be relaxed because they based it on the $1\sigma$ confidence level (68\%) only and neglected the fact that the mass and radius measurements were correlated.
	K
	.~Chatziioannou and S.~Han~\cite{Chatziioannou:2019yko} used two different hadronic EoS, namely 
	DBHF
	and SFHo
	together with the same CSS model EoS to find out by Bayesian methods that the onset mass and the strength of a strong first order phase transition could be constrained with 50 - 100 neutron star merger detections.
	
	Miller et al.~\cite{Miller:2019nzo} based their study on a multi-polytrope EoS to assess the impact of the improvement of the precision on the determination of the
	nuclear symmetry energy in the laboratory for the determination of the compact star EoS, particularly below nuclear saturation densities. 
	An important aspect that is implemented (following~\cite{Alvarez-Castillo:2016oln}) and stressed in this paper 
	is the need for a variable lower limit on the maximum star mass since it is an observational input for the Bayesian analysis and thus subject to updates.
	A hard lower bound implementation leads to the loss of important information since this value changes with every new observation, albeit the uncertainties 
	inherent in the measurements themselves; see~\cite{Raaijmakers:2018bln}, which is also based on a multi-polytrope parameterization. 
	Moreover, a follow-up work by Miller et al.~\cite{Miller:2019cac} extended the EoS to include first-order phase transitions in the multi-polytrope approach.
	
	Among the several state-of-the-art studies is the work of Most et al.~\cite{Most:2018hfd}, which considered a large set of EoS parametrizations with and without phase transitions, fitted to the multi-polytrope parametrization. 
	They imposed constraints on the upper bound for the maximum compact star mass to be less than 2.16 $M_{\odot}$, as well as on the tidal
	deformability to be less than 800. 
	The resulting $R$ value for a 1.4$M_{\odot}$ pure hadronic star lies in the range $12.00~{\rm km}< R_{1.4} < 13.45~{\rm km}$ and for a hybrid twin star within $8.53~{\rm km}< R_{1.4} < 13.74~{\rm km}$ at the 2$\sigma$ confidence level (95\%). 
	Interestingly, other recent works found consistent results. 
	The follow-up study of Raaijmakers et al.~\cite{Raaijmakers:2019dks} featured two EoS classes: 
	a five-polytrope piecewise EoS and a piecewise constant speed of sound EoS together with the measurements of the GW170817 event, the NICER data for 
	PSR J0030+0451, the most massive NS of about 2.14$M_{\odot}$, and chiral effective field theory (EFT) constraints. 
	Capano et al.~\cite{Capano:2019eae} used a speed of sound parameterization based on 
	chiral EFT constraints, as well, but different priors to the ones used by Raaijmakers et al., resulting in skewed radius results in lower values for the radii
	$9.2~{\rm km}< R_{1.4} < 12.3~{\rm km}$, similar to the value found by Lim et al.~\cite{Lim:2019som}, who based their studies on empirical nuclear observables.
	
	Studies that are based on analyzing the density dependence of the speed of sound and using the new multi-messenger data are conclusive about the 
	possibility to violate the conformal bound $c_s^2 \le 1/3$ in compact stars \cite{Fujimoto:2019hxv,Tews:2018kmu,Reed:2019ezm,Zdunik:2012dj}.
	This fact alone, however, does not yet allow concluding about the composition of neutron star interiors since such a violation occurs under neutron star conditions for standard nuclear matter EoS \cite{Reed:2019ezm}, as well as in CFL quark matter \cite{Zdunik:2012dj}, where typically $c_s^2\approx 0.6$.
	Disregarding this fact, the authors of \cite{Annala:2019puf} made the misleading statement that high-density quark matter in compact stars has to obey the conformal bound and drew conclusions from it.
	
	Essick et al.~\cite{Essick:2019ldf} used a non-parametric EoS Bayesian inference under different priors, also finding consistent results.
	The LIGO-Virgo collaboration presented a study that incorporated realistic compact star equations of state under the framework of
	long-lived rotation remnants following the compact star merger~\cite{LIGOScientific:2019eut}. 
	The result was that in the case of a slowly rotating remnant (less than 1.9 kHz), the baryonic mass of the
	static compact star turned out to be 3.5$M_{\odot}$, allowing for ruling out three of the EoS they considered.
	Last but not least, in~\cite{Traversi:2020aaa}, for the first time, the results from X-ray observations of accretion-powered pulsars in low-mass X-ray binaries 
	were included in the Bayesian analysis together with the LIGO-Virgo Collaboration (LVC) 
	results for GW170817 and the NICER results for PSR J0030+0451 to find based on a relativistic mean-field EoS that $R_{1.4}\sim 12$ km. 
	These authors reported that in a version of the model that included $\Lambda$ hyperons, the radius should be $R_{1.4}\sim 14$ km, at tension with the constraints from GW170817. 
	
	It is this glimpse of a tension between the GW170817 data favoring smaller radii and the NICER, as well as X-ray pulsar data that nourishes our work on EoS with a strong phase transition that could accommodate a bimodality of radius measurements for similar values of the NS mass, the mass twin phenomenon 
	\cite{Blaschke:2013ana,Benic:2014jia,Alvarez-Castillo:2014dva,Bejger:2016emu,Alvarez-Castillo:2016oln,Christian:2017jni,Alvarez-Castillo:2017qki,Kaltenborn:2017hus,Ayriyan:2017nby,Paschalidis:2017qmb,Alford:2017qgh,Montana:2018bkb,Sieniawska:2018zzj,Alvarez-Castillo:2018pve,Maslov:2018ghi,Li:2019fqe,Bozzola:2019tit,Blaschke:2019tbh,Christian:2019qer,Jakobus:2020nxw}
	obeying the constraint from the existence of $2~M_\odot$ pulsars. 
	
	\subsection{Goals of the Present Work}
	We investigate the question of how likely it may be that GW170817 was not the merger of two canonical neutron stars, but involved at least one if not two hybrid stars with a quark matter core.
	To this end, we develop a Bayesian analysis method for selecting the most probable equation of state (EoS) under a set of constraints from compact star physics, 
	which now include the tidal deformability from GW170817 and the first result for the mass and radius determination for PSR J0030+0451 by the NICER Collaboration.
	We apply this method for the first time to a two-parameter family of hybrid EoS that is based on realistic models for hadronic matter (DD2 with nucleonic excluded volume) and color superconducting quark matter (generalized nonlocal NJL (nlNJL) model), which produce a third family of hybrid stars in the $M-R$ diagram.
	One parameter ($\mu_<$) characterizes the chemical potential for the onset of quark deconfinement, while the other 
	($\Delta_P$) belongs to the mixed phase construction that mimics the thermodynamics of pasta structures and includes the Maxwell construction for 
	$\Delta_P=0$, as well as the Glendenning construction \cite{Glendenning:1992vb} for $\Delta_P\sim 6\%$ \cite{Maslov:2018ghi} as limiting cases.
	As has been demonstrated, e.g., in \cite{Bhattacharyya:2009fg}, the type of mixed phase construction has an influence on the stability of compact star sequences.
	Many general classes of EoS like multi-polytropes are presently used for Bayesian analyses with multi-messenger observational data from neutron stars. 
	The specific choice in the present work of a class of hybrid EoS that describes mass twin solutions in a wide range of masses with their actual position depending on the parameter $\mu_<$ is motivated by their unique capability to resolve a possible tension between radius measurements in the same mass 
	range for GW170817 (e.g., $R=11.0^{+0.9}_{-0.6}$ km \cite{Capano:2019eae}) and for PSR J0030+0451 ($R=13.02^{+1.24}_{-1.06}$ km \cite{Miller:2019cac}).
	
	This paper is organized as follows. 
	The next section contains a description of the equation of state followed by a section describing the methods of computing the neutron star configurations and tidal deformabilities. 
	Then follows a section describing our Bayesian analysis that goes beyond our previous work \cite{Ayriyan:2018blj} by now including the first results of the mass and radius determination for the millisecond pulsar PSR J0030+0451 of the NICER team \cite{Miller:2019cac}, including the results of the actual measurements, as well as a fictitious result as one possible outcome of a forthcoming NICER analysis.
	In the final section, we present our conclusions and an outline of ongoing research. 
	
	\section{Quark-Hadron Hybrid Equation of State }
	
	The hybrid equation of state is the basis of our work and consists of three ingredients. 
	The Skyrme model EoS SLy4 \cite{Douchin:2000kx,Haensel:2004nu,Pearson:2012hz} is employed for the crust in the subsaturation region and matched to the density-dependent 
	relativistic mean-field (RMF) EoS ``DD2'' \cite{Typel:2009sy} describing the properties of atomic nuclei and nucleonic matter around the saturation density with stiffening effects of a nucleonic excluded volume of 4 fm$^3$ according to Typel \cite{Typel:2016srf} acting at supersaturation densities only. 
	The nomenclature for this EoS is ``DD2p40''. 
	In a recent work \cite{Blaschke:2020qrs} it has been demonstrated that the stiffening effect of this EoS is quantitatively justified by the Pauli blocking among nucleons due to their quark substructure.
	For the high-density phase of matter, we use the microscopic quark matter EoS, which is matched to the hadronic phase by a mixed phase construction.
	The quark matter EoS and the mixed phase construction are both described in the following subsections.
	
	\subsection{Quark Matter Model}
	The quark matter approach to the EoS consists of the nonlocal NJL model \cite{Blaschke:2007ri} for which a generalization was introduced in 
	\cite{Alvarez-Castillo:2018pve}, which consisted of adopting a chemical potential dependence of the vector mean-field coupling $\eta(\mu)$ 
	and a chemical potential-dependent bag pressure $B(\mu)$ {by}:
	\begin{equation}
	\label{eq:twofold}
	P(\mu) = P(\mu;\eta(\mu),B(\mu))= P_{\rm nlNJL} (\mu;\eta(\mu)) - B(\mu)~,
	\end{equation}
	where the density-dependent vector coupling is given by: 
	\begin{equation}
	\label{eq:eta-mu}
	\eta(\mu)=\eta_> f_\gg(\mu)+ \eta_< f_\ll(\mu)
	\end{equation} 
	and the density-dependent bag function is:
	\begin{equation}
	\label{eq:bag}
	B(\mu)=Bf_<(\mu)f_\ll(\mu)~.
	\end{equation} 
	Here, two smooth switch functions are introduced, one that changes from one to zero at a lower chemical potential $\mu_<$ with a width $\Gamma_<$, 
	\begin{equation}
	\label{eq:f<}
	f_{<}(\mu)=\frac{1}{2}\left[1-\tanh\left(\frac{\mu-\mu_<}{\Gamma_<}\right)\right], 
	\end{equation}
	and one that switches off at a higher chemical potential $\mu_\ll$ with a width $\Gamma_\ll$, 
	\begin{equation}
	\label{eq:f<<}
	f_{\ll}(\mu)=\frac{1}{2}\left[1-\tanh\left(\frac{\mu-\mu_\ll}{\Gamma_\ll}\right)\right], 
	\end{equation}
	together with the complementary functions that serve to switch on,
	\begin{equation}
	f_{>}(\mu) = 1 - f_<(\mu)~,~~f_{\gg}(\mu) = 1 - f_{\ll}(\mu). 
	\end{equation}
	With these switch functions, the density-dependence is introduced in such a way that the additional parameters have an intuitively clear meaning. 
	The vector coupling switches from a value $\eta_< $ at low chemical potentials to $\eta_> $ at high ones, whereby the switch is positioned at $\mu_\ll$ and extends over a region of the width $\Gamma_\ll$. For the introduction of the chemical potential-dependent vector coupling to obtain heavy hybrid stars, see also \cite{Blaschke:2013ana}. 
	This density dependence of the vector coupling facilitates the density-dependent stiffening of quark matter as a prerequisite for obtaining a third family of compact stars and the directly related mass twin phenomenon. For this, two conditions must be simultaneously fulfilled: 
	(1) a strong deconfinement transition with a sufficiently large jump of the energy density so that it induces a gravitational instability (requiring a relatively soft quark matter EoS at low densities) and 
	(2) a high maximum mass of the hybrid star sequence to fulfill the $2~M_\odot$ constraint (requiring a stiff quark matter EoS at higher densities). 
	We note that an alternative to the introduction of a density-dependent vector meson coupling would be the use of a higher order vector interaction like the eight quark interaction of~\cite{Benic:2014oba} with constant couplings for which it has been demonstrated in \cite{Benic:2014jia} that both the above conditions can be fulfilled.
	
	The bag pressure switches in a region of width $\Gamma_<$ around $\mu= \mu_<$ from a maximal value at low chemical potentials to zero at high ones, with a slight modification by the factor $f_\ll(\mu)$. This bag function and its behavior are motivated in the following way. 
	In dynamical chiral quark models like the nlNJL, there is a mean-field pressure associated with the dynamical chiral symmetry breaking by the appearance of a chiral (quark) condensate. 
	It is well known that this contribution to the thermodynamics of the quark model can be interpreted as a bag pressure 
	(see, e.g., \cite{Buballa:1996tm,Buballa:2003qv}) with a chemical potential dependence that was extracted in \cite{Grigorian:2003vi} within a color superconducting nonlocal NJL model under neutron star conditions. 
	While this effect of the melting of the quark condensate at increasing density is inherent in the dynamical quark model, one can expect that a similar contribution to the thermodynamics should come from the melting of the gluon condensate. 
	As the gluon dynamics is not explicit in the nlNJL model, it may therefore be in order to superimpose a density-dependent bag pressure that should resemble nonperturbative effects from the gluon sector, which are to cease at the deconfinement transition \cite{Blaschke:2010vj}. 
	In the present work, we use the parameter $\mu_<$ in order to vary the onset density of the deconfinement transition, while the other parameters 
	remain fixed at the values of Set 2 from~\cite{Alvarez-Castillo:2018pve}.
	
	\subsection{Mixed Phase Construction}
	
	For the completion of the compact star EoS, the hadron and quark matter EoS are joined either by a Maxwell construction or an interpolation procedure
	aimed at mimicking the effect of pasta phases that may exist at the interface. 
	As for the latter case, we employ the replacement interpolation method (RIM) \cite{Ayriyan:2017tvl,Abgaryan:2018gqp} 
	that replaces an EoS region around the Maxwell construction point by a polynomial function of the type:
	\begin{equation}
	P_{M}\left(\mu\right)=\alpha_{2}\left(\mu-\mu_{c}\right)^{2} +
	\alpha_{1}\left(\mu-\mu_{c}\right)+\left(1+\Delta_{P}\right)P_{c} \, ,
	\end{equation}
	with $\alpha_{1}$, $\alpha_{2}$, $\mu_{H}$, and
	$\mu_{Q}$, which are determined from the continuity
	conditions at the hadron and quark border points $\mu_{H}$ and $\mu_{Q}$,
	\begin{eqnarray}
	P_{H}\left(\mu_{H}\right)&=&P_{M}\left(\mu_{H}\right) \, , \\
	P_{Q}\left(\mu_{Q}\right)&=&P_{M}\left(\mu_{Q}\right)\, , \\
	n_{H}\left(\mu_{H}\right)&=&n_{M}\left(\mu_{H}\right)\, , \\
	n_{Q}\left(\mu_{Q}\right)&=&n_{M}\left(\mu_{Q}\right).
	\end{eqnarray}
	The resulting EoS connects the three points $P_{H}(\mu_{H})$, $P_{c}+\Delta P=P_{c}(1+\Delta_{P})$, and $P_{Q}(\mu_{Q})$. 
	The parameter $\Delta P$ corresponds to an extra pressure at the Maxwell construction point; thus, we use it as a free parameter to quantify the effect of 
	the mixed phase of the hadron to quark transition.
	
	\subsection{Constant Speed of Sound Extrapolation}
	
	Due to a backbending of the given quark curves on the $P$-$\varepsilon$ plot, a causality violation at high energy densities appears. 
	To avoid such a violation, the constant speed of sound (CSS) extrapolation was implemented for the quark tables. 
	The CSS EoS for this extrapolation is given by
	
	\begin{eqnarray}
	P(\mu) & = & P_0 + P_1 \left({\mu}/{\mu_x} \right)^\beta , \hspace{2cm} \mathrm{for~} \mu>\mu_x,\\
	\varepsilon(\mu) & = &-P_0 + P_1 (\beta -1) \left({\mu}/{\mu_x} \right)^\beta ,\hspace{6mm}  \mathrm{for~} \mu>\mu_x,\\
	n_B(\mu) & = & P_1 \displaystyle \frac{\beta}{\mu_x} \left({\mu}/{\mu_x} \right)^{\beta-1}, \hspace{2cm}  \mathrm{for~} \mu>\mu_x,
	\label{eq:ccs}
	\end{eqnarray}
	
	where $\mu_x$ is the chemical potential where the matching to the CSS extrapolation starts. 
	The parameter $\beta$ is directly related to the squared speed of sound: 
	\begin{equation}
	c_s^2=\frac{\partial P/ \partial \mu}{\partial \varepsilon / \partial \mu} = \frac{1}{\beta - 1}~,
	\label{eq:css}
	\end{equation}
	so that
	\begin{equation}
	\beta = 1 + \displaystyle \frac{1}{c_s^2} ~.
	\label{eq:css_beta}
	\end{equation}
	It is obvious that fulfilling the causality constraint $c_s^2 \le 1$ entails that $\beta \ge 2$. 
	The coefficients $P_0$ and $P_1$ in Eq. (\ref{eq:ccs}) are defined as:
	\begin{eqnarray}
	P_0 & = & \left[(\beta - 1) P_x - \varepsilon_x \right]/\beta\\
	P_1 & = & \left(P_x + \varepsilon_x \right) / \beta~,
	\label{eq:css_pp}
	\end{eqnarray}
	where $P_x=P(\mu_x)$ and $\varepsilon_x=\varepsilon(\mu_x)$.
	
	The results below have been obtained for the choice of the following rule.
	For those EoS for which $c_s^2$ has a peak (below $c_s^2=1$) at an energy 
	density beyond the upper limit $\varepsilon_Q$ of the deconfinement 
	phase transition, the squared speed of sound stays constant at the value
	it has at $\varepsilon_x=818$ MeV/fm$^3$.
	For the other EoS, the squared speed of sound stays at $c_s^2=1$ 
	for $\varepsilon>\varepsilon_x$, where $\varepsilon_x$ is the energy density
	where it first reached this value.
	In the latter case of the CSS extrapolation with $c_s^2 = 1$, the parameter 
	$\beta=2$ for all chemical potentials $\mu>\mu_x$ where $\mu_x$ is
	defined as $\varepsilon(\mu_x)=\varepsilon_x$. 
	
\begin{figure}[H]
		\centering
		\vspace{-5mm}
		\includegraphics[width=0.48\textwidth]{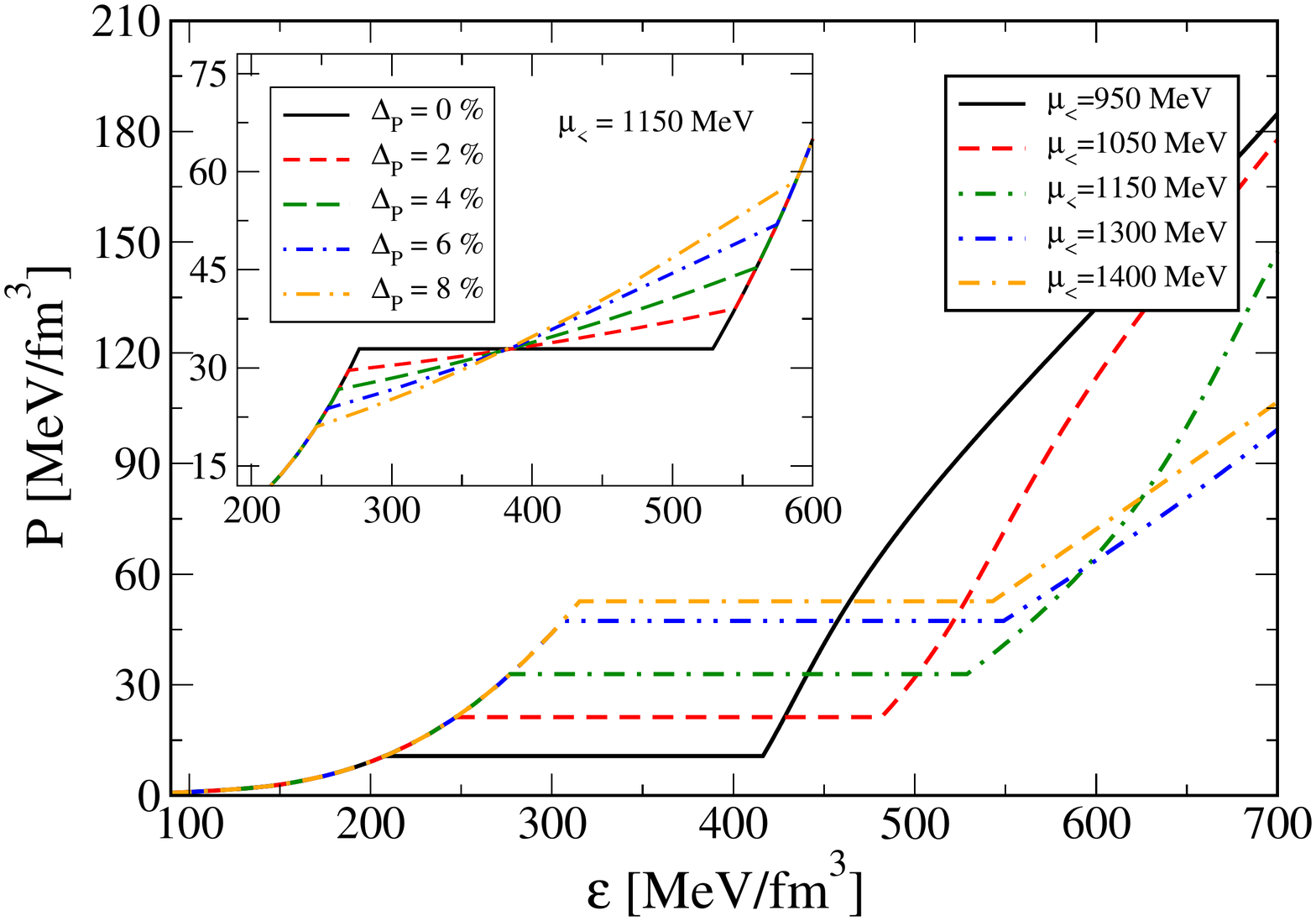} 
		\includegraphics[width=0.48\textwidth]{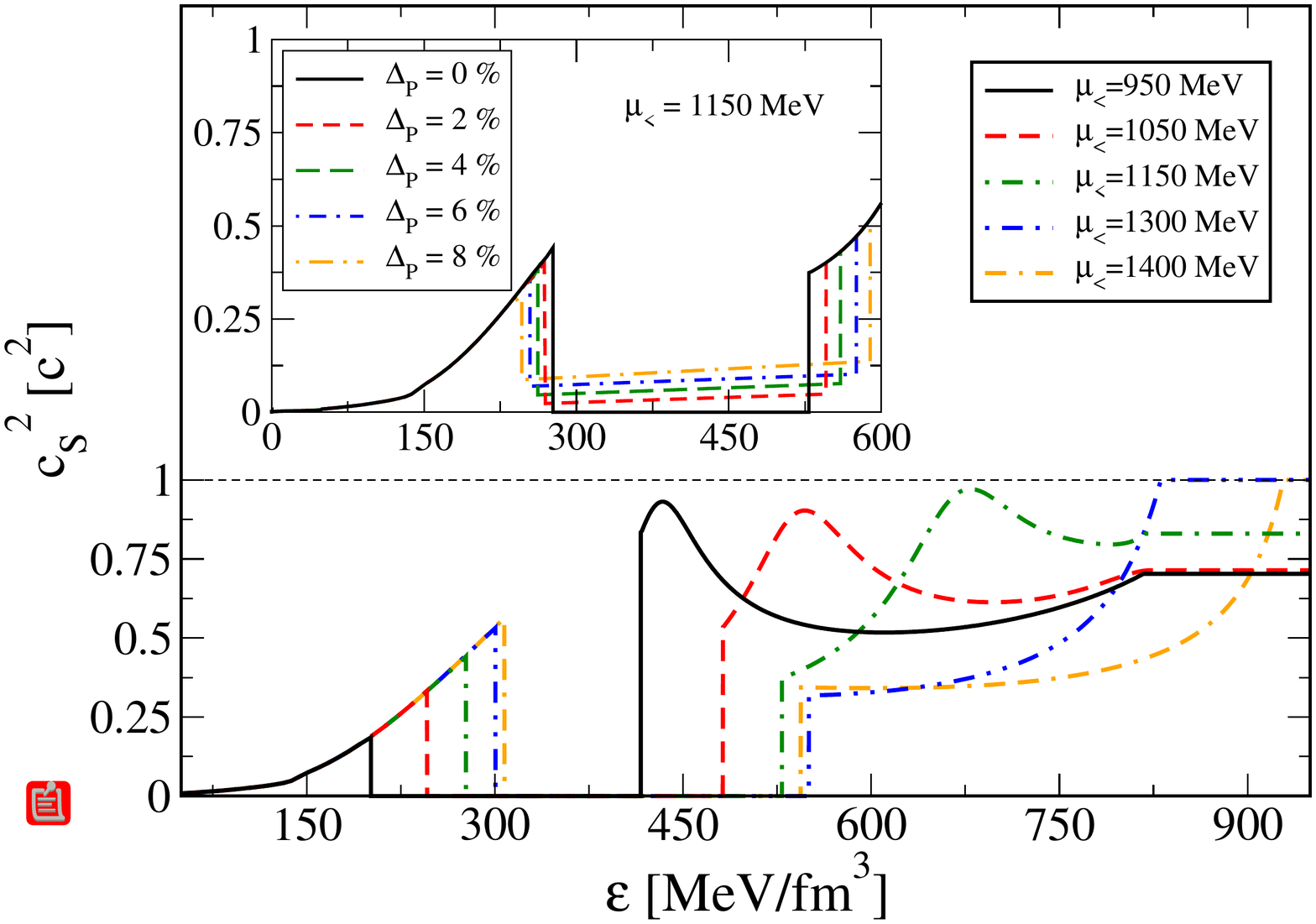} 
		\caption{\label{fig:eos}
			Left panel: 
			Hybrid star equation of state with the hadronic branch from the RMF model DD2p40 and the quark matter branch from the 
			nonlocal chiral quark model nlNJL with different onset of deconfinement parametrized by $\mu_<=950, 1050, 1175, 1300, 1400$ MeV. 
			For the phase transition, the Maxwell construction is used. At large energy density, 
			$\varepsilon>818$ MeV/fm$^3$, the EoS is matched with a constant speed of sound (CSS) model. 
			Right panel: Squared speed of sound as a function of the energy density for the EoS of the left panel. 
			The inset figures on both panels show the mixed phase construction with $\Delta_P=0 (2) 8$ \% for the case $\mu_<=1150$ MeV.                                                                                                                                                                                                       
		} 
	\end{figure}

	The resulting EoS $P(\varepsilon)$ and the squared speed of sound $c_s^2(\varepsilon)$ are 
	shown in the left and right panels of Figure ~\ref{fig:eos}, respectively.

	\section{Compact Star Configurations}
	
	In this section, we describe how we compute the sequences of hybrid compact stars and their tidal deformabilities, which have now become accessible to observation with the gravitational wave detectors of the LIGO-Virgo Collaboration and the analysis of neutron star merger events like GW170817.
	
	\subsection{Mass and Radius}
	
	Using the full compact star EoS, we obtain the mass-radius relations for stellar sequences by solving the Tolman--Oppenheimer--Volkoff (TOV) equations, which describe a static, spherical star:
	\begin{eqnarray}
	\label{eq:TOVa}
	\frac{dP( r)}{dr}&=& 
	-\frac{\left(\varepsilon( r)+P( r)\right)
		\left(m( r)+ 4\pi r^3 P( r)\right)}{r\left(r- 2m( r)\right)},\\
	\frac{dm( r)}{dr}&=& 4\pi r^2 \varepsilon( r)~,
	\label{eq:TOVb}
	\end{eqnarray}
	where $P(r)$ is the pressure profile, related to the energy density profile $\varepsilon(r)$ by the EoS.
	The latter defines the profile function $m(r)$ of the mass enclosed up to the distance $r$ from the center of the spherical star configuration. 
	The boundary conditions $P(r=R)=0$, $M=m(r=R)$ determine the total stellar mass $M$ and radius $R$, respectively, in dependence of the initial value 
	$P_c= P(r=0)$ for the integration of the TOV equations. 
	By varying this initial value, or equivalently the central energy density $\varepsilon_c = \varepsilon(r=0)=\varepsilon(P_c)$, one obtains the $M-R$ 
	relationship, which is uniquely characterizing the EoS. 
	This equivalence between the EoS of compact star matter and the $M-R$ relation for the compact star sequences it generates is the basis for constraining 
	the EoS by mass and radius measurements on pulsars.
	One of the landmarks of a sequence of stars in the $M-R$ diagram is the maximum mass, which delimits a branch (family) of them that is stable against gravitational collapse and can thus be subject to comparison with observations.
	The observation of a pulsar mass that exceeds the maximum mass can be used to exclude the corresponding EoS.

	\begin{figure}[H]
		\centering
		\includegraphics[width=0.7\textwidth]{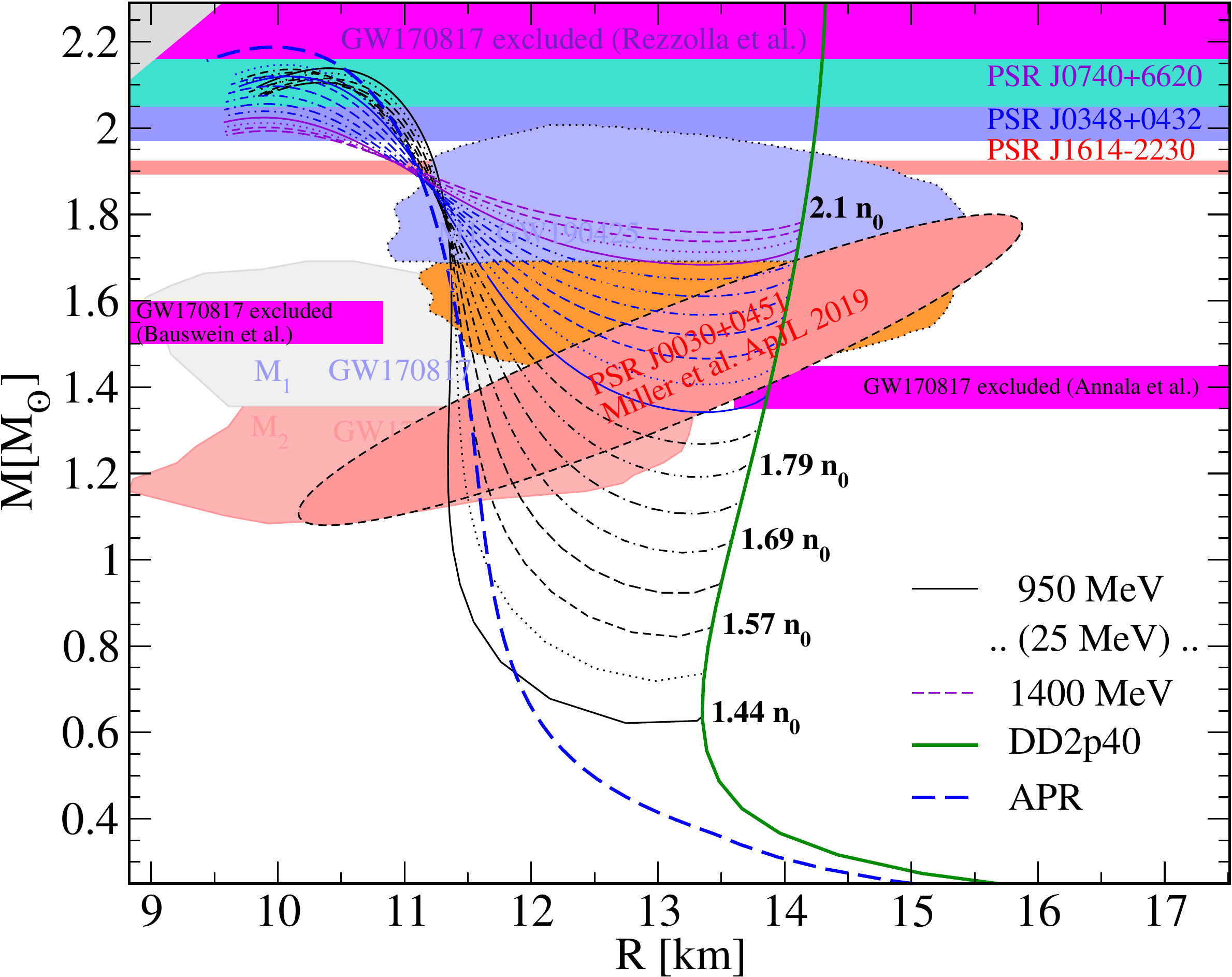}
		\caption{\label{fig:M-R}
			Mass-radius diagram with solutions of the Tolman--Oppenheimer--Volkoff (TOV) equations for the EoS with a Maxwell construction ($\Delta_P=0$) of the deconfinement transition.
			Solutions for $\Delta_P\neq 0$ are shown below in Figures~\ref{fig:MR-PEpost}, \ref{fig:MR-PEpost2}, and \ref{fig:MR-PEpost3}.
			For comparison, the modern constraints from multi-messenger astronomy are shown: the mass of 
			PSR J0740+6620 \cite{Cromartie:2019kug},
			the compactness constraint on tidal deformabilities \cite{Abbott:2018exr} obtained by an analysis of the gravitational wave signal from the inspiral phase of the binary compact star merger GW170817, and one of both the mass and radius measurement for PSR J0030+0451 by NICER
			\cite{Miller:2019cac}.
			Shown also for an orientation is the mass-radius constraint reported in~\cite{Abbott:2020uma} for the binary compact star merger GW190425, but it is not used in the Bayesian analysis presented here.
			Next to the selected points of onset of the deconfinement transition, we indicate the corresponding baryon number densities in units of the saturation density $n_0=0.15$ fm$^{-3}$.
		} 
	\end{figure}
	
	In Figure ~\ref{fig:M-R}, the $M-R$ relations for the class of hybrid EoS introduced in the previous section (see also Figure ~\ref{fig:eos}) are shown for different values of the parameter $\mu_< = 950~(25)~1400$ MeV characterizing the onset of deconfinement and restricting us here to the case 
	of the Maxwell construction ($\Delta_P=0$) of the phase transition. The $M-R$ relations including mixed phase constructions with $\Delta_P\neq 0$ are shown and discussed below in Figures~\ref{fig:MR-PEpost}, \ref{fig:MR-PEpost2}, and \ref{fig:MR-PEpost3}.
	We note that this class of hybrid EoS has a sufficiently strong phase transition in order to generate ``twins'' in the $M-R$ diagram: For a given gravitational mass 
	$M$ in a certain narrow range, there exist two possible star configurations with different radii on disconnected stable branches, belonging to the second family of ordinary neutron stars (here represented by the DD2p40 EoS) and to the third family of hybrid stars. 
	This feature is quite robust against mixed phase effects \cite{Ayriyan:2017nby}. 
	
	We would like to note that our hybrid star solutions exhibit the feature of a special point (narrow region) through which all $M(R)$ curves pass.
	This feature of TOV solutions has been analyzed using analytic methods by Yudin et al.~\cite{Yudin:2014mla} where they found that it should be related to the (approximate) constant speed of sound in the quark matter phase. 
	It is particularly pronounced for hybrid EoS with strong phase transitions that exhibit third family solutions and the mass twin star phenomenon as, e.g., in
	\cite{Kaltenborn:2017hus,Blaschke:2010vd,Alvarez-Castillo:2018pve},
	even if the speed of sound is not strictly constant.
	
	The unstable branch between the second and third family is characterized by a positive slope. We did not remove it here in Figure ~\ref{fig:M-R} in order to guide the eye from the third family branch to the onset of the phase transition, where the unstable branch joins the second family. 
	We also indicate next to a few points of the phase transition onset the corresponding baryon density in units of the saturation density $n_0=0.15$ fm$^{-3}$, which range from $1.44\, n_0$ to $2.10\, n_0$, which are typical densities for switching from nuclear theories to models of high-density matter \cite{Tews:2019cap}.
	
	When the phase transition proceeds through a mixed phase (when $\Delta_P\neq 0$), then the mass $M_{\rm onset}$ at the onset of the transition is lowered as compared to the Maxwell-construction case, and eventually, the disconnected branches join so that there are no longer any twins.
	
	All hybrid star EoS in this study fulfill the maximum mass constraint that is represented here by the Shapiro-delay based mass measurement on PSR J0740+6620
	\cite{Cromartie:2019kug}. Former observational maximum mass constraints from PSR J0348+0432 \cite{Antoniadis:2013pzd} and PSR J1614-2230 
	\cite{Demorest:2010bx,Fonseca:2016tux,Arzoumanian:2017puf} are also shown, but no longer as constraining. 
	The first joint observation of a gravitational wave signal from the inspiral phase of the merger GW170817 and the related electromagnetic signals in different bands of the spectrum (the kilonova) resulted in additional constraints shown as magenta shaded excluded regions in Figure ~\ref{fig:M-R}. 
	The constraint on the compactness of pulsars itself, which was derived from the measurement of the tidal deformability by the LIGO-Virgo Collaboration 
	\cite{Abbott:2018exr}, is shown as a shaded heart-shaped region in the mass range between $1.16$ and $1.60~M_\odot$. 
	Shown also for an orientation is the mass-radius constraint reported in~\cite{Abbott:2020uma} for the binary compact star merger GW190425, but it is not used in the Bayesian analysis presented in the following section. The cigar-shaped region in Figure ~\ref{fig:M-R} is one of both mass and radius measurements for PSR J0030+0451 by NICER \cite{Miller:2019cac}. The second one \cite{Riley:2019yda} is sufficiently similar and shall be explicitly considered in a subsequent work.
	For a comparison, the soft hadronic EoS by Akmal, Pandharipande, and Ravenhall (APR) \cite{Akmal:1998cf} is also shown. 
	We note that in the relevant mass range for GW170817 where tidal deformability constraints are available, the curves for the hybrid EoS with low 
	$M_{\rm onset}<M_\odot$ are almost indistinguishable from that for APR. 
	
	\subsection{Tidal Deformabilities and GW170817}
	
	The tidal deformability (TD) for a given star configuration is computed following the approach based on a perturbation of the static metric of the star \cite{Hinderer:2009ca}. 
	The resulting perturbation functions, together with the TOV mass and pressure profiles, are the ingredients for the determination of the Love number $k_2$ \cite{Hinderer:2007mb}. 
	The dimensionless TD is defined as $\Lambda=\lambda/M^{5}$ in terms of the deformability $\lambda$ related to the Love number:
	\begin{equation}
	k_{2}=\frac{3}{2}\lambda R^{-5}.
	\end{equation}
	
\begin{figure}[H]
		\centering
		\vspace{-5mm}
		\includegraphics[width=0.48\textwidth]{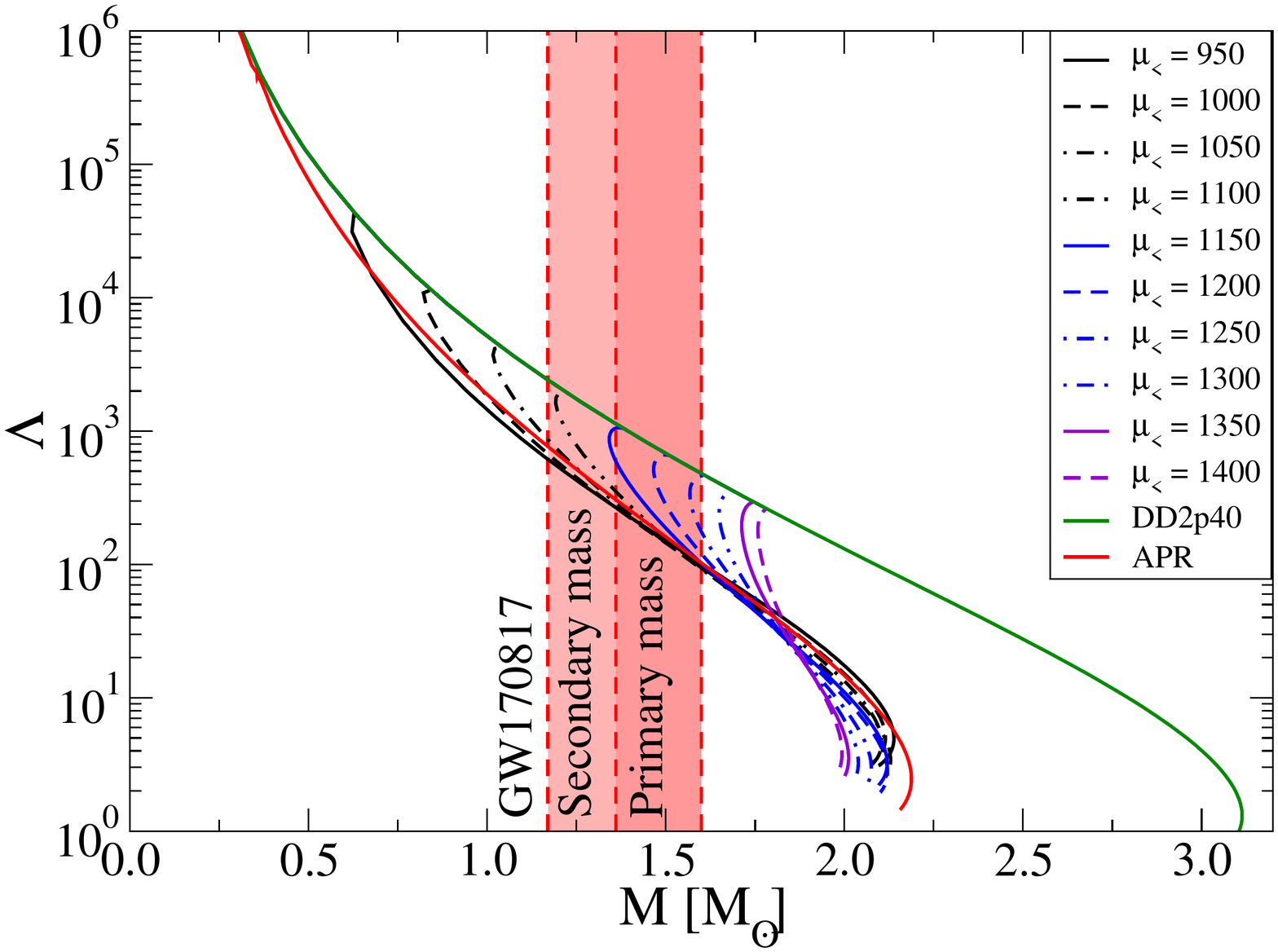} 
		\includegraphics[width=0.48\textwidth]{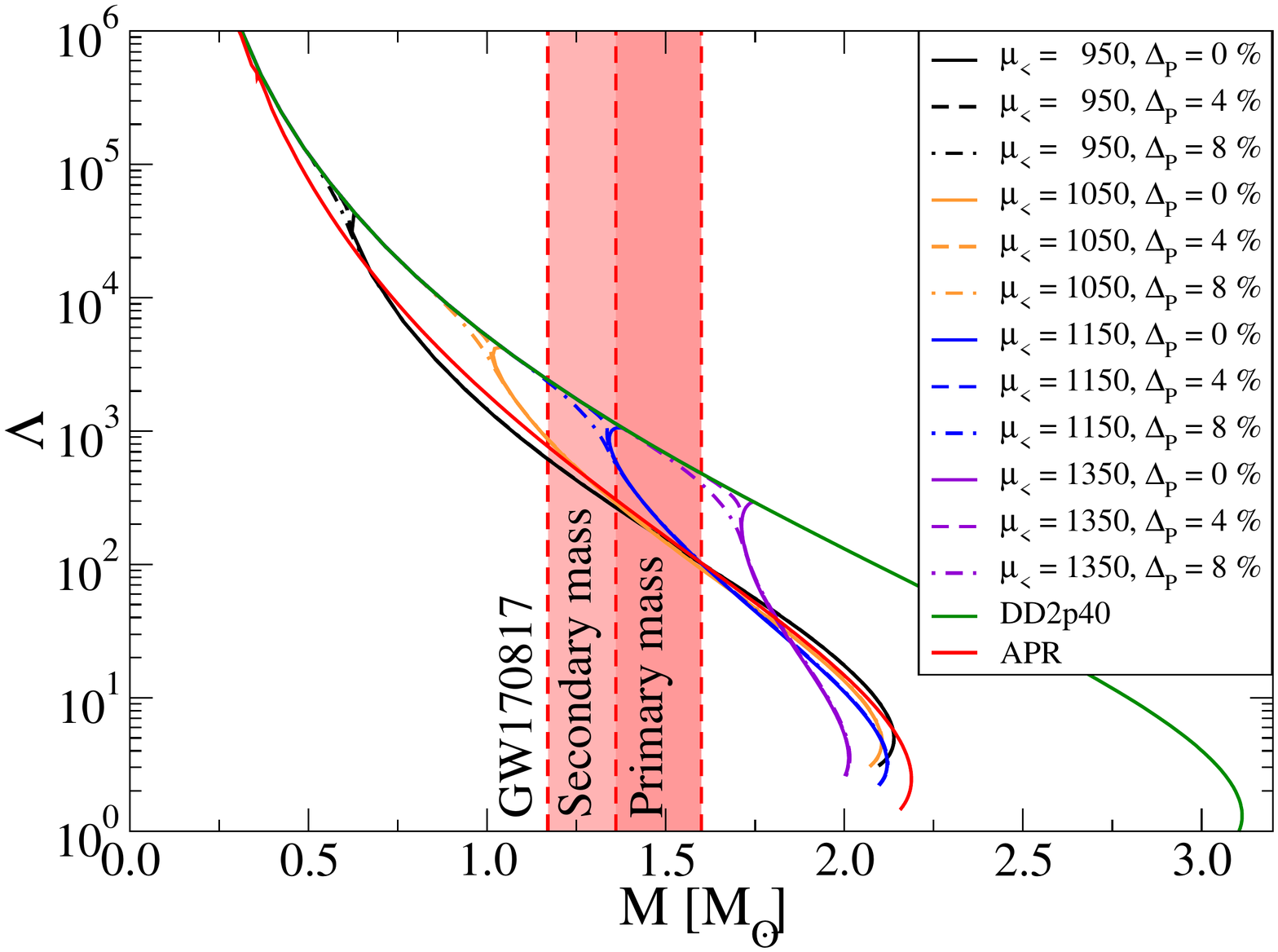} 
		\caption{\label{fig:L-M}
			Tidal deformability as a function of the gravitational mass $M$ of the star. Vertical dashed lines indicate the lower ($1.16~M_\odot$) and upper ($1.60~M_\odot$) limits of the low-spin prior merger masses, as well as the equal mass case ($M_1=M_2=1.365~M_\odot$) for GW170817 \cite{TheLIGOScientific:2017qsa}. 
			In the left panel, we show a subset of hybrid EoS for selected values of $\mu_<$ obtained by Maxwell construction ($\Delta_P=0$). In the right panel for the four values of $\mu_<=950$, 1050, 1150, and 1350 MeV, we show the effect of structures in the mixed phase mimicked by the construction with nonvanishing 
			parameter $\Delta_P=4$ and $8\%$. 
			For comparison, in both panels, results for the soft hadronic Akmal, Pandharipande, and Ravenhall (APR) EoS are shown, which in the mass range probed by the merger GW170817 are almost indistinguishable from those of the hybrid EoS with early onset $\mu_< \le 1050$ MeV. 
		} 
	\end{figure}

	In Figure ~\ref{fig:L-M}, we show the TD as a function of the gravitational mass $M$ of the star. The vertical dashed lines indicate the lower ($1.16~M_\odot$) and upper ($1.60~M_\odot$) limits of the low-spin prior merger masses, as well as the equal mass case ($M_1=M_2=1.365~M_\odot$) for GW170817 \cite{TheLIGOScientific:2017qsa}. 
	In the left panel, we show a subset of hybrid EoS, which were obtained by a Maxwell construction ($\Delta_P=0$) for selected values of $\mu_< = 950~(50)~1400$ MeV. 
	In the right panel, we show the effect of structures in the mixed phase for each of the four values of $\mu_<=950$, 1050, 1150, and 1350 MeV by choosing the mixed phase parameter $\Delta_P=0,~4\%$, and $8\%$. 
	For comparison, in both panels, the TD for the soft hadronic APR EoS is shown, which in the mass range probed by the merger GW170817 is almost indistinguishable from those of the hybrid EoS with early onset $\mu_< \le 1050$ MeV. 
	
	\begin{figure}[H]
		\centering
		\vspace{-5mm}
		\hspace{-5mm}
		\includegraphics[width=0.52\textwidth]{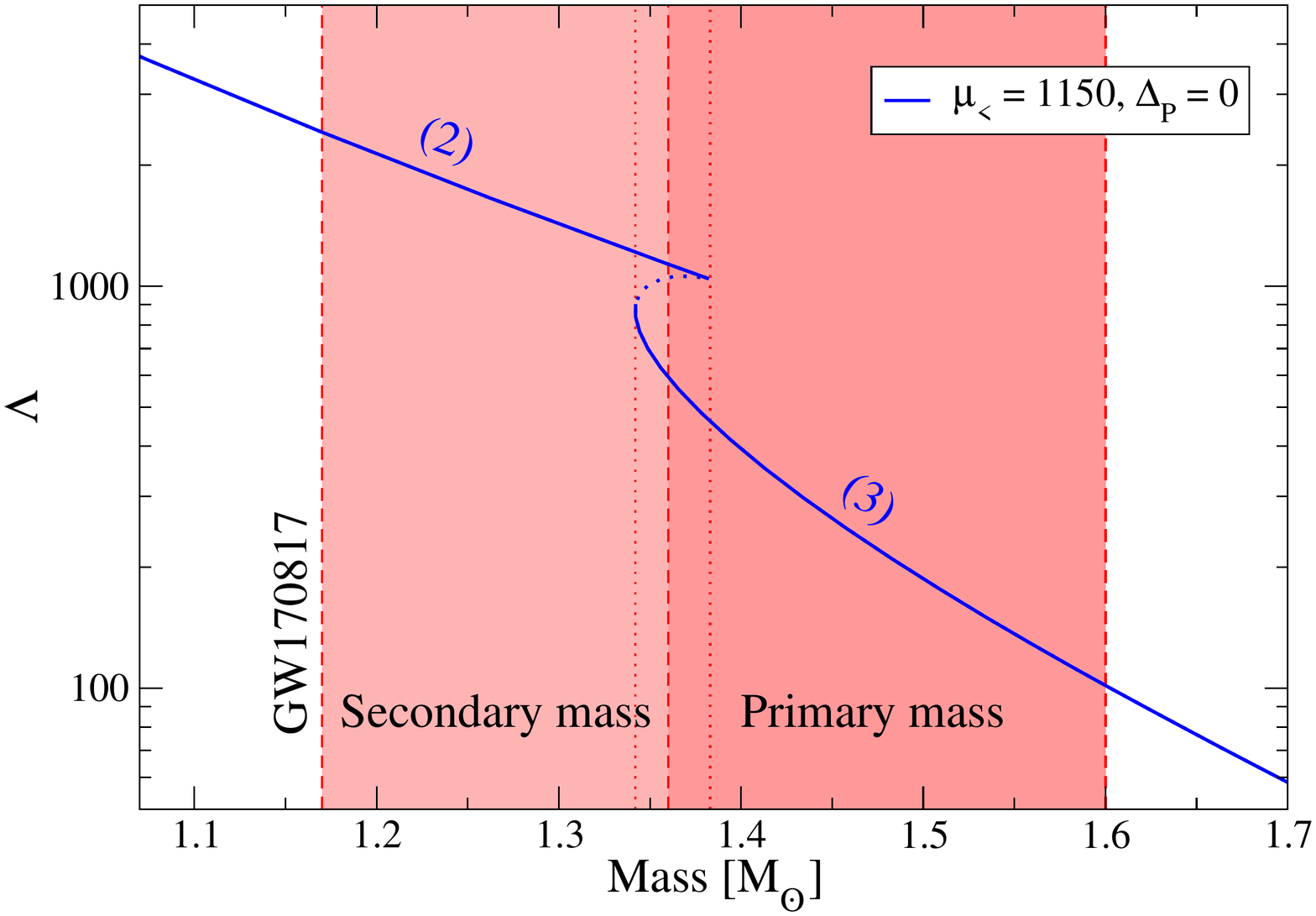} \hspace{-5mm}
		\includegraphics[width=0.52\textwidth]{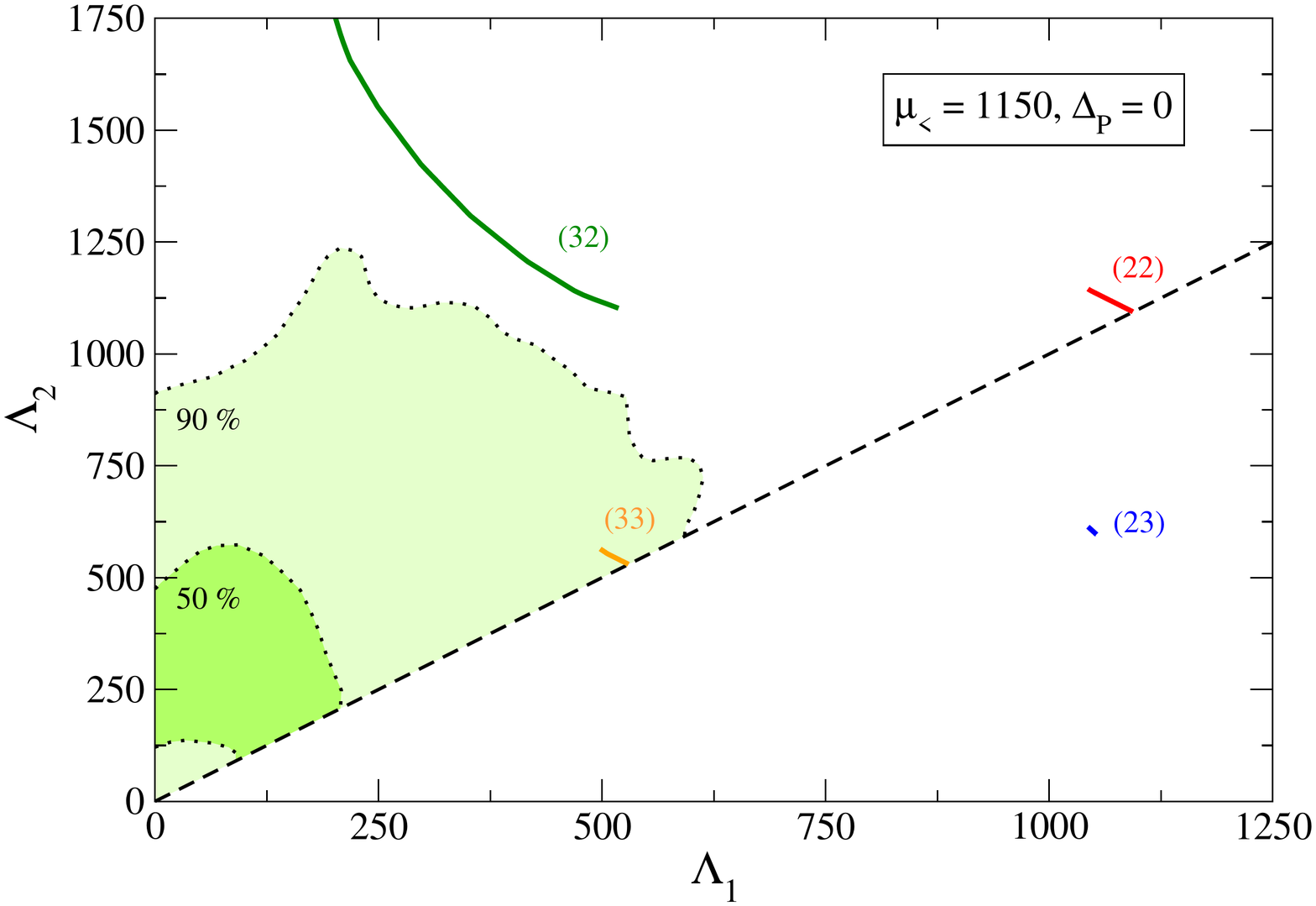} 
		\caption{\label{fig:LL1150}
			The tidal deformability constraint for the special case of the Maxwell construction ($\Delta=0$) of the hybrid EoS with $\mu_<=1150$ MeV when the stars in the merger GW170817 could belong to both second and third family branches (left panel), so that one out of four regions in the $\Lambda_{1} - \Lambda_{2}$ diagram can be populated for this one case of the EoS (right panel). 
		} 
	\end{figure}
	
	In Figure ~\ref{fig:LL1150}, we examine the particular case of the parameter set $\mu_<=1150$ MeV and $\Delta_P=0$ for which the equal-mass case of GW170817 with $M_1=M_2=1.36~M_\odot$ lies in the middle of the twin mass range $1.342~M_\odot<M<1.383~M_\odot$ for this parametrization of the hybrid EoS. 
	In the left panel, we show the dimensionless tidal deformability $\Lambda$ as a function of the gravitational mass $M$ of the star and indicate the mass ranges from which the primary and the secondary mass in the binary can be chosen for the low-spin prior case. Within the twin mass range (vertical dotted lines), two values of $\Lambda$ are possible for the same $M$, lying on the second and third family branch of compact stars, labeled by (2) and (3), resp.
	There are four combinations of stars possible in the binary GW170817, which fulfill the constraint of the measured total mass 
	$M_{\rm tot}=2.73^{+0.04}_{-0.01}~M_\odot$ and which we label with two numbers as (22), (23), (33), and (32), denoting the origin of the primary and secondary star, resp., stemming from the second (2) or third (3) family. With the twofold degeneracy in the case of mass twins, there are four lines generated when the tidal deformabilities $\Lambda_1$ and $\Lambda_2$ of the two stars in a binary are plotted against each other. 
	These lines have endpoints that form the corners of a square in the $\Lambda_1 - \Lambda_2$ diagram shown in the right panel of Figure ~\ref{fig:LL1150}.
	These endpoints correspond to the equal-mass case. 
	Of course, any specific merger event like GW170817 is represented by just one point in the $\Lambda_1 - \Lambda_2$ diagram so that different merger events that cover the same mass interval and have overlap with the twin mass range may result in different peaks in the probability landscape on the $\Lambda_1 - \Lambda_2$ plane. 
	
	We note that for the case of compact star mass triples, which were introduced in \cite{Alford:2017qgh} and discussed in the context of GW170817 in 
	\cite{Li:2019fqe}, the mass-degenerate case would correspond to nine points, lying on the corners of a 3 $\times$ 3 quadratic array in the $\Lambda_1 - \Lambda_2$ diagram. We would like to draw attention to the fact that EoS with mass twins and triples are examples for which the $\Lambda_1 - \Lambda_2$ diagram 
	is not reflection symmetric to the diagonal $\Lambda_1 = \Lambda_2$ because of the existence of separate families of stars corresponding to one and the same EoS. This situation appears as if the two components in the binary that are lying on different star branches were described by different EoS, and it should be taken into consideration in the analysis of the gravitational wave signals. 
	If the true compact star EoS would allow mass twin (or triple) stars in a mass range accessible to merger events, then in a combined analysis of the inspiral GW signals from many events in that very mass interval should emerge a multi-peaked (four or nine peaked) probability landscape in the $\Lambda_1 - \Lambda_2$ plane.
	
	\begin{figure}[H]
		\centering
		\vspace{-5mm}
		\hspace{-5mm}
		\includegraphics[width=0.52\textwidth]{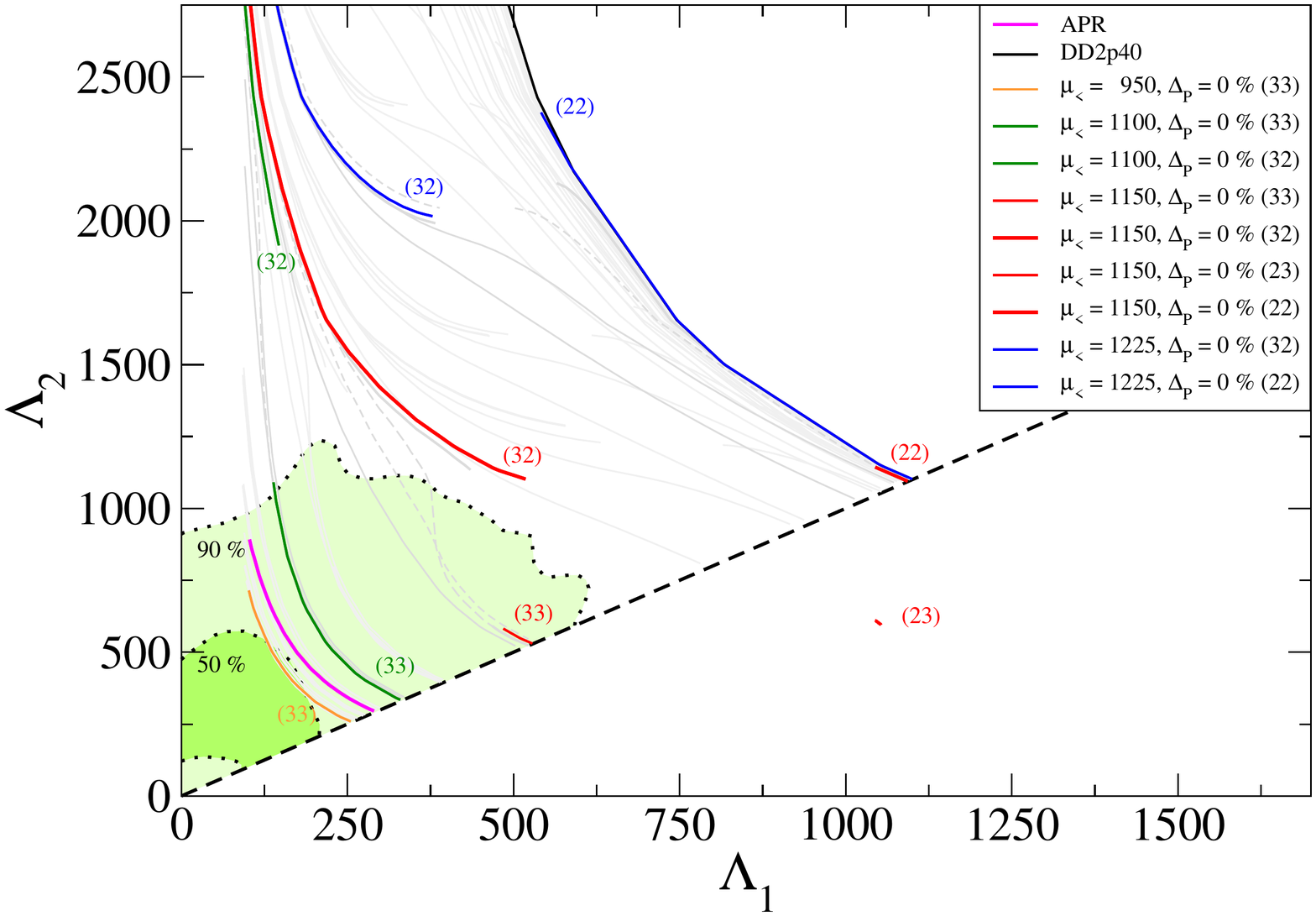} \hspace{-5mm}
		\includegraphics[width=0.52\textwidth]{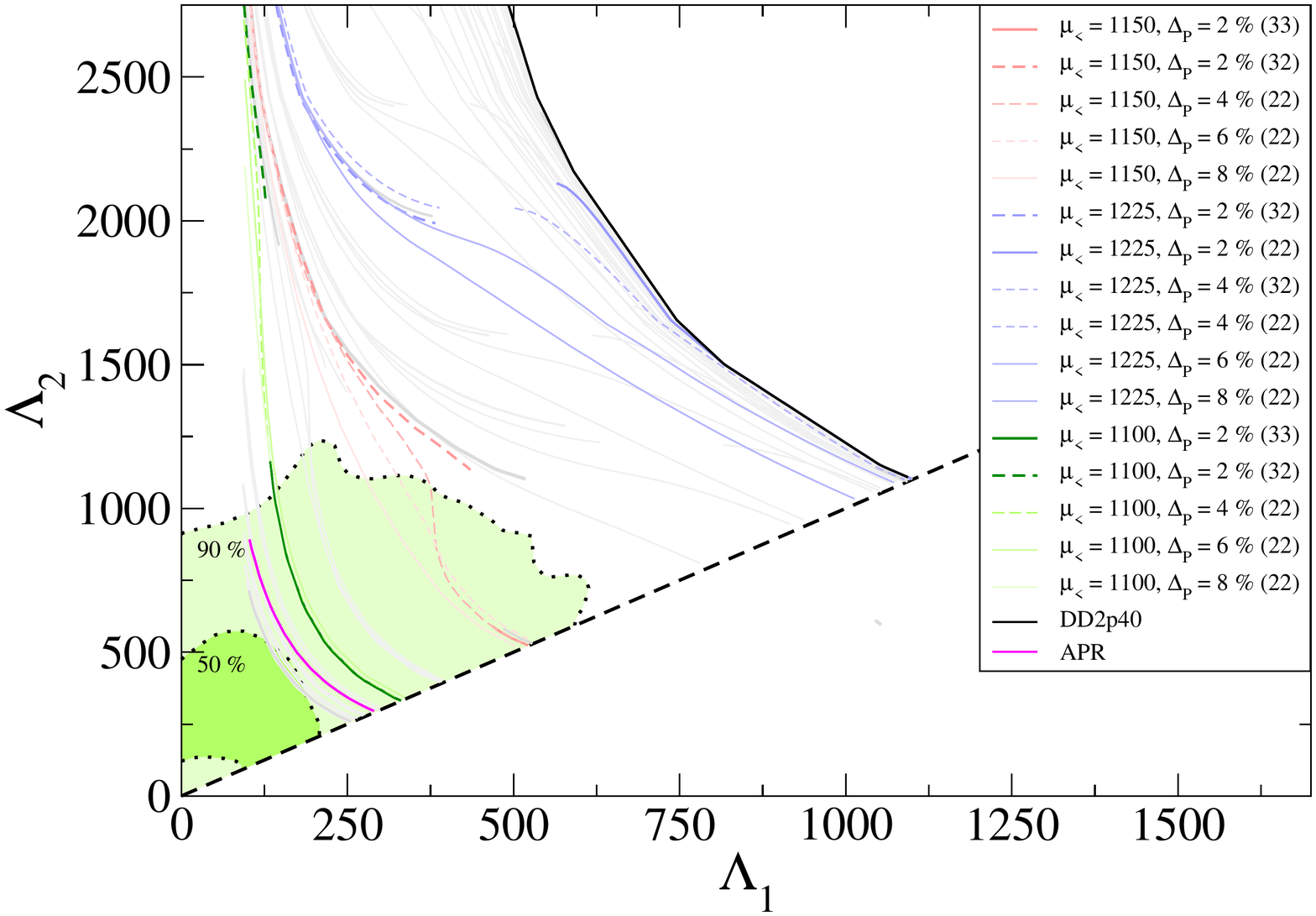} 
		\caption{\label{fig:L1-L2}
			Tidal deformability constraint from the binary compact star merger GW170817 in the $\Lambda_1-\Lambda_2$ diagram. 
			According to the LVC
			analysis \cite{Abbott:2018exr}, the light and dark green regions correspond to 90\% and 50\% confidence, respectively; see also \cite{LIGO}.
			In the left panel, we display results for the hybrid EoS with selected values (all in MeV) of the parameter $\mu_<$ for the onset of deconfinement in the Maxwell construction case;
			see also Figures~\ref{fig:L-M} and~\ref{fig:LL1150}. The right panel shows the effect of the mixed phase construction for these selected values of $\mu_<$ and 
			$\Delta_P=2$, 4, 6, and 8\% highlighted in color.
		} 
	\end{figure}
	
	In Figure ~\ref{fig:L1-L2}, we explore how the pattern in the $\Lambda_1 - \Lambda_2$ plane would evolve when the mass ranges covered by the twin phenomenon and by the merger would be shifted against each other. Here, we consider the latter fixed at that of GW170817 and vary the former by changing $\mu_<$, the parameter for the onset of deconfinement in the EoS; in the left panel for the Maxwell transition case ($\Delta_P=0$) and in the right panel for different mixed phase parameters $\Delta_P=2, ~4,~6$, and 8\%.
	For the transition at low densities ($\mu_< < 1100$ MeV) and thus low $M_{\rm onset}<M_\odot$, the merger probes the hybrid star branch only, which is almost indistinguishable from that for the soft APR EoS. For high transition densities ($\mu_< > 1250$ MeV), the onset mass exceeds the merger mass range, and thus, the tidal deformabilities are those of the stiff hadronic EoS DD2p40.
	In the intermediate range $1100~{\rm MeV}<\mu_< < 1250~{\rm MeV}$, as a consequence of the overlap between merger and twin mass ranges, multiple disconnected branches occur in the $\Lambda_1 - \Lambda_2$ diagram. Besides the binary merger of two second family stars (22) and that of two third family stars (33), also the mixed cases occur where a pure neutron star from the second family merges with a hybrid star from the third family (23) and (32).
	In the right panel of Figure ~\ref{fig:L1-L2}, one can follow how the separate branches would merge to a single line when the mixed phase parameter increases, 
	and this would reflect the joining of the second and third family branches for the twin EoS in the $M-R$ diagram.
	An observational tomography of the $\Lambda_1 - \Lambda_2$ plane equivalent to the theoretical one presented here would require the detection of GW signals from a large number of sufficiently ``loud'' merger events that would allow for a precise measurement of tidal deformabilities and to map different mass windows in order to ``detect'' the emergence of disconnected maxima in the probability landscape as a signal for the mass twin phenomenon.

	\section{Bayesian Inference for the EoS Models}
	
	\subsection{Vector of Parameters}
	The set of parameters defining a set of models could be represented in the parameter space by introducing the vector of parameters. 
	Each vector then corresponds to a particular model defined by the value of $\mu_<$ related to the onset density of deconfinement and the value $\Delta_P$ 
	specifying the transition construction,
	\begin{equation}
	\label{pi_vec}
	\overrightarrow{\pi}_i = \left\{{\mu_{<}}_{(j)},{\Delta_P}_{(k)}\right\},
	\end{equation}
	where $i = 0 (1) N-1$ and $i = N_2\times j + k$,~$j = 0 (1) N_1-1$,~$k = 0 (1) N_2-1$ and $N_1$ and $N_2$ are number of values of model parameters 
	$\mu_{<}$ and $\Delta_P$, respectively.
	
	\subsection{Likelihood of a Model under the $\Lambda_1$--$\Lambda_2$ Constraint from GW170817 }
	
	In order to implement the tidal deformability constraint on the compact star EoS, reflected on the $\Lambda_1$--$\Lambda_2$ diagram that includes probability regions from the GW170817 event~\cite{TheLIGOScientific:2017qsa,Abbott:2018exr}, we employ for the likelihood the formula:
	\begin{eqnarray}
	\label{eq:lhoodLL}
	P\left(E_{GW}\left|\pi_i\right.\right) = \int_{l_{22}} \beta(\Lambda_1(\tau), \Lambda_2(\tau))d\tau + \int_{l_{23}} \beta(\Lambda_1(\tau), \Lambda_2(\tau))d\tau
	\nonumber\\
	+ \int_{l_{32}} \beta(\Lambda_1(\tau), \Lambda_2(\tau))d\tau + \int_{l_{33}} \beta(\Lambda_1(\tau), \Lambda_2(\tau))d\tau, 
	\end{eqnarray}
	where $l_{ps}$ is a line in the $\Lambda_1$-$\Lambda_2$ plane formed by all admissible combinations of masses for the primary ($p$) and the secondary ($s$) star in the binary GW170817. 
	The probability distribution function (PDF) $\beta(\Lambda_1, \Lambda_2)$ is given for GW170817 and the free parameter $\tau$, which, for instance, is the central density of the primary star, included in Equation \eqref{eq:lhoodLL}.
	The PDF $\beta(\Lambda_1, \Lambda_2)$ (see~Figure ~\ref{fig:L1-L2}) was reconstructed (as previously in~\cite{Ayriyan:2018blj}) by the Gaussian kernel density estimation method with $\Lambda_1$-$\Lambda_2$ data given on the LIGO web-page \cite{LIGO}.
	
	\subsection{Likelihood of a Model under the Constraint on the Lower Limit of the Maximum Mass}
	
	The likelihood of the model is the conditional probability of the expected value of the possible maximum mass for the given model parameter vector: 
	\begin{equation}
	\label{eq:lhoodMass}
	P\left(E_{A}\left|\pi_i\right.\right) = \Phi(M_i, \mu_A, \sigma_A)= \frac{1}{2}\left[1+ {\rm erf} \left(\frac{M_i-\mu_A}{\sqrt{2}\sigma_A} \right) \right],
	\end{equation}
	where $M_i$ is the maximum mass of the model given by $\pi_i$. 
	The values for $\mu_A = 2.14~\mathrm{M_{\odot}}$ and $\sigma_A = 0.095~\mathrm{M_{\odot}}$ are taken from the mass measurement of PSR J0740+6620, $M=2.14_{-0.09}^{+0.10}~\mathrm{M_{\odot}}$~\cite{Cromartie:2019kug}. 
	We note that we combined this new measurement with the previously used mass measurement for the two solar-mass pulsar J0348+0432 with 
	$M=2.01_{-0.04}^{+0.04}~\mathrm{M_{\odot}}$~\cite{Antoniadis:2013pzd} as a product of independent likelihoods and thus obtained a strengthening of the selective power of the maximum mass constraint because in this way, the models with a lower maximum mass get additionally suppressed.

	\subsection{Likelihood of a Model under the Combined $M$-$R$ Constraint of the NICER Experiment}
	
\begin{figure}[H]
		\centering
		\includegraphics[width=0.7\textwidth]{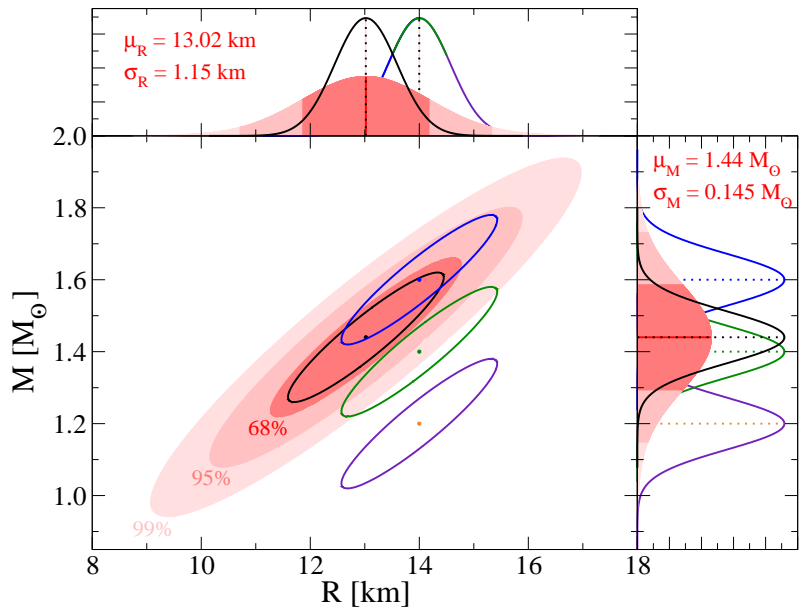}
		\caption{\label{fig:NICER}
			Probability measure for the mass-radius measurement of the NICER experiment for PSR J0030+0451 using values from the analysis of~\cite{Miller:2019cac} (reddish areas). 
			The colored lines correspond to the $2\sigma$-regions (95\% confidence areas) for four fictitious mass and radius measurements for which the error ellipse is narrowed by a factor of two relative to the actual NICER measurement, while the expectation values for mass and radius are unchanged (black lines), and the radius is shifted to 14 km and the mass to 1.6 $M_\odot$ (blue lines), 1.4 $M_\odot$ (green lines), and 1.2 $M_\odot$ (orange lines).		} 
	\end{figure}

	We implemented in our Bayesian analysis the first simultaneous $M$-$R$ measurement as provided by the NICER experiment for the millisecond pulsar PSR J0030+0451 \cite{Riley:2019yda,Miller:2019cac}.
	We represent this measurement by a bivariate normal distribution in the $M$-$R$ diagram: 
	
	\begin{equation}
	\label{eq:bivariate}
	\mathcal{N}(M,\,R;\,\mu_M,\,\sigma_M,\,\mu_R,\,\sigma_R,\,\rho)=\displaystyle \frac{1}{2 \pi \sigma_M \sigma_R\sqrt{1-\rho^2}}\exp\left(-\frac{x}{2(1-\rho^2)}\right)
	\end{equation}
	with:
	\begin{equation}
	x=\displaystyle \frac{(M-\mu_M)^2}{\sigma_M^2}-2\rho \displaystyle \frac{(M-\mu_M)(R-\mu_R)}{\sigma_M\sigma_R}+\displaystyle \frac{(R-\mu_R)^2}{\sigma_R^2},
	\end{equation}
	where the parameters
	$\mu_M = 1.44\, {M}_\odot$, $\sigma_M = 0.145\, {M}_\odot$, $\mu_R = 13.02\,\textrm{km}$, $\sigma_R = 1.150\,\textrm{km}$, 
	and the correlation coefficient $\rho = 0.9217$ were calculated within the approximation (\ref{eq:bivariate}) to the result of \cite{Miller:2019cac}.
	This corresponds to a tilted ellipse; see the red areas in Figure ~\ref{fig:NICER}. 
	The likelihood for a model $\pi_i$ under the constraint from the NICER experiment is:
	\begin{eqnarray}
	\label{eq:lhoodLL}
	P\left(E_{MR}\left|\pi_i\right.\right) = \int_{l_{2}} \mathcal{N}(\mu_R,\,\sigma_R,\,\mu_M,\,\sigma_M,\rho)d\tau
	+ \int_{l_{3}} \mathcal{N}(\mu_R,\,\sigma_R,\,\mu_M,\,\sigma_M,\rho)d\tau,
	\end{eqnarray}
	where the lines for the second and third families in the $M$-$R$ diagram are denoted as $l_2$ and $l_3$, respectively. 
	
	Besides the real NICER measurement, one can assume fictitious $M-R$ measurements that could be a possible future result for either the same object or 
	another one. 
	We explore the consequences for the Bayesian analysis when such a measurement for four cases shown by colored lines in Figure ~\ref{fig:NICER}
	would have a standard deviation 
	reduced by a factor two to $\sigma_M = 0.0725\, {M}_\odot$, $\sigma_R = 0.575\,\textrm{km}$, and the mean values of mass and radius would be (i) unchanged (black lines), (ii) $\mu_R = 14\,\textrm{km}$, $\mu_M = 1.6~{M}_\odot$
	(blue lines), (iii) $\mu_R = 14\,\textrm{km}$, $\mu_M = 1.4~{M}_\odot$ (green lines), and (iv) $\mu_R = 14\,\textrm{km}$, $\mu_M = 1.6~{M}_\odot$ (orange lines).
	We varied the possible outcome of the mass measurement within the limits of the mass range of the merging neutron stars in GW170817. 
	
	The reason for considering fictitious measurements with the mean value of the radius beyond the limits posed by the tidal deformability analysis of GW170817 is that we
	can demonstrate the discovery potential of the class of EoS with the third family and twin star solutions we consider here. 
	For this class of microphysically motivated EoS, it would be no contradiction to observe significantly different radii in the same mass range; just the opposite. Such an observation would be supporting the mass twin case and its underlying physics: a sufficiently strong phase transition with a large 
	latent heat fulfilling the Seidov criterion \cite{Seidov:1971,Alford:2013aca,Alvarez-Castillo:2017qki}. 
	We focus on this case because we are convinced that this possible outcome of a future NICER mass and radius measurement would be a breakthrough finding in the field with far-reaching consequences for the phenomenology of strong interactions, the EoS, and the quest for the critical point in the QCD phase diagram
	\cite{Blaschke:2013ana}.
	Postulating a fictitious radius measurement with a smaller radius than the 13.02 km reported in \cite{Miller:2019cac}, for instance 12 km in the GW170817 mass range, would then overlap with the constraint derived from the LVC analysis \cite{Abbott:2018exr}; see also Figure ~\ref{fig:M-R}.
	Therefore, even an increase in the accuracy of the NICER measurement by lowering the variances $\sigma_M$ and $\sigma_R$ has a negligible effect on the 
	outcome of the Bayesian analysis as we checked. In that case, the low onset mass solutions would be strongly favored, as in the case of the actual NICER measurement with larger variance, or when the Bayesian analysis would be performed without including the NICER result; see the middle panel of Figure ~\ref{fig:BA}. 
	
	\subsection{Posterior Distribution}
	The full likelihood for the given $\pi_i$ can be calculated as a product of all likelihoods, since the considered constraints are conditionally independent of 
	each other:
	\begin{equation}
	\label{eq:p_event}
	P\left(E\left|\overrightarrow{\pi}_{i}\right.\right)= \prod_{m} P\left(E_{m}\left|\overrightarrow{\pi}_{i}\right.\right),
	\end{equation}
	where $m$ is the index of the constraints.
	The posterior distribution of models on the parameter diagram is given by Bayes' theorem:
	\begin{equation}
	\label{eq:bayes}
	P\left(\overrightarrow{\pi}_{i}\left|E\right.\right)=\frac{P\left(E\left|\overrightarrow{\pi}_{i}\right.\right)P\left(\overrightarrow{\pi}_{i}\right)}{\sum\limits _{j=0}^{N-1}P\left(E\left|\overrightarrow{\pi}_{j}\right.\right)P\left(\overrightarrow{\pi}_{j}\right)},
	\end{equation}
	where $P\left(\overrightarrow{\pi}_{j}\right)$ is the prior probability distribution in the space of N models that is taken to be uniform, 
	$P\left(\overrightarrow{\pi}_{j}\right)=1/N$. 
	
	\subsection{Results of the Bayesian Analysis}
	
\begin{figure}[H]
		\centering
		\includegraphics[width=0.32\textwidth]{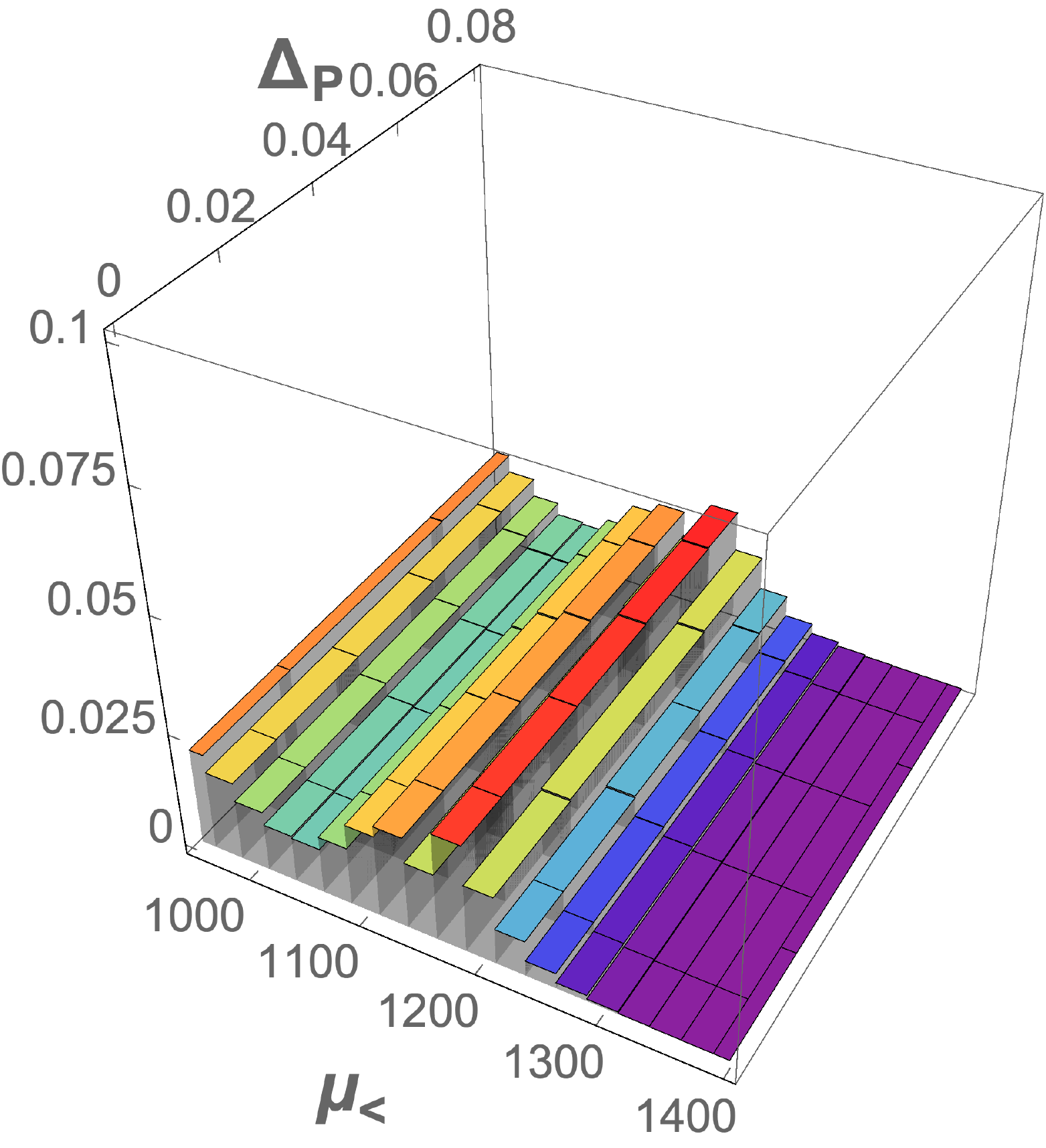} 
		\includegraphics[width=0.32\textwidth]{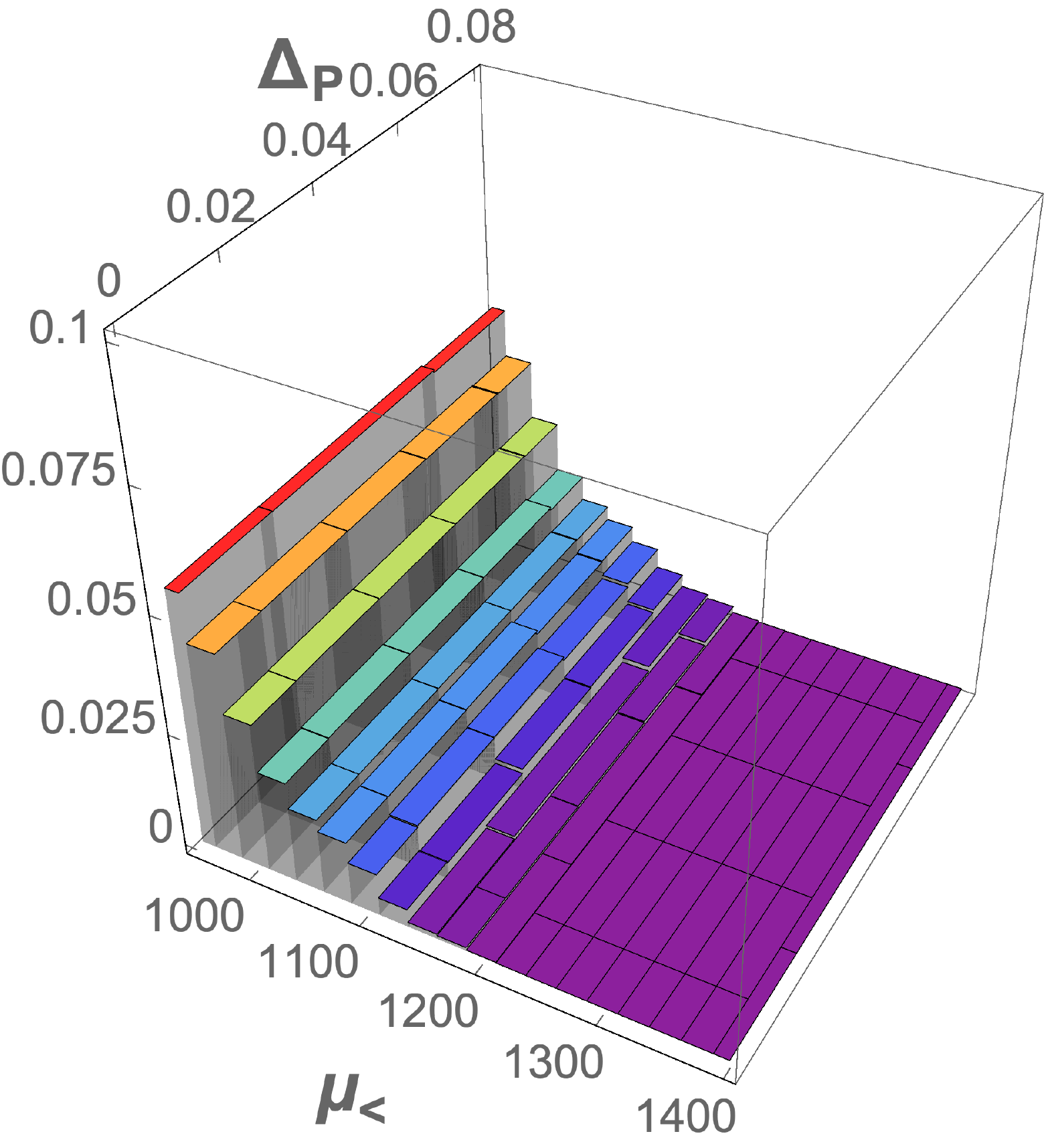} 
		\includegraphics[width=0.32\textwidth]{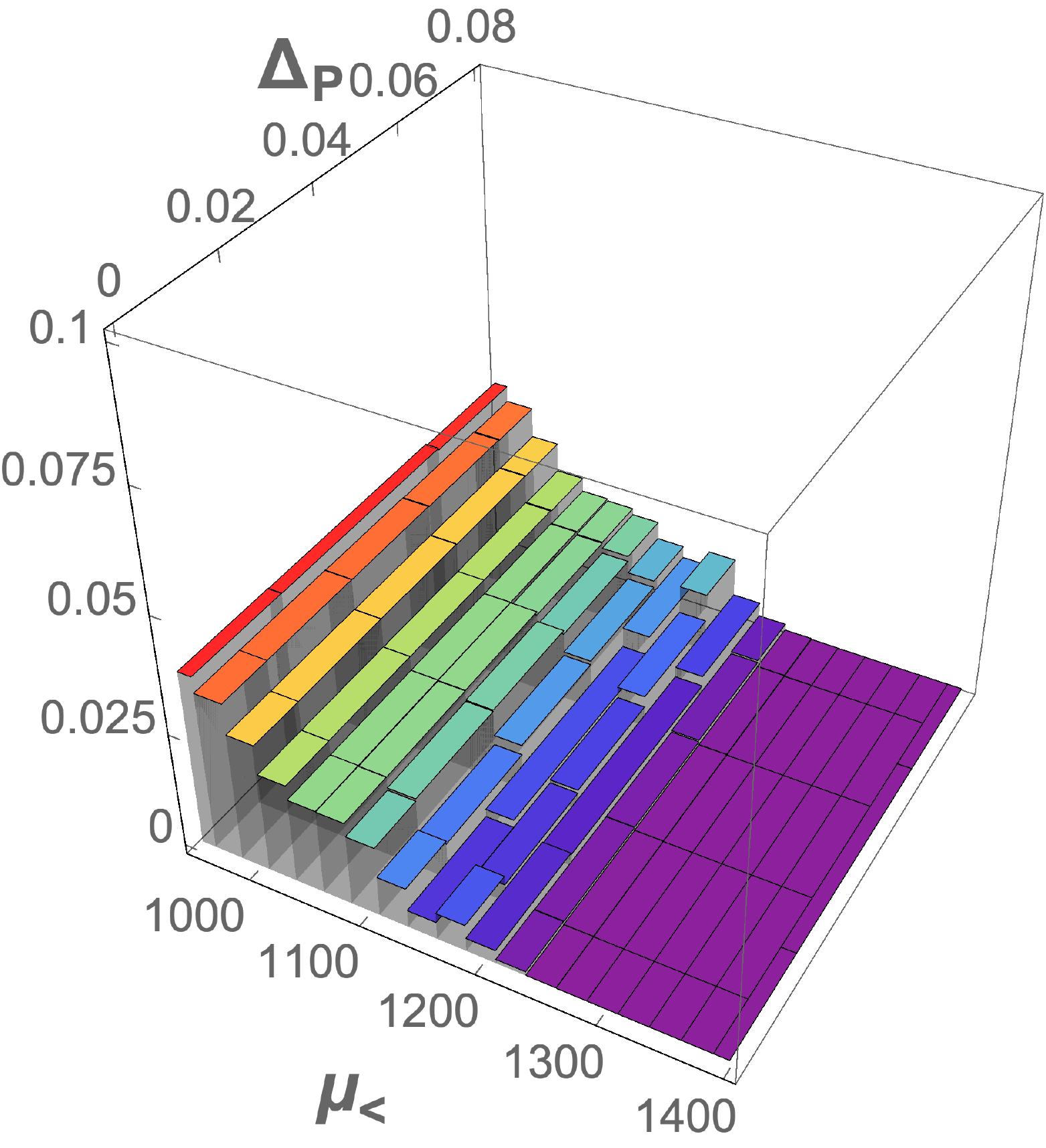} 
		\caption{\label{fig:BA}
			Posterior probabilities for different models after Bayesian analysis using only the mass measurements of PSR J0740+6620 \cite{Cromartie:2019kug} and PSR J0348+0432 \cite{Antoniadis:2013pzd} as a lower limit on the maximum mass (left panel); this mass constraint together with the compactness constraint from tidal deformabilities of the binary compact star merger GW170817 \cite{Abbott:2018exr} (middle panel) and the constraint from one of the mass-radius measurement by NICER \cite{Miller:2019cac} (right panel). 
			In all three cases, the probability is indifferent to changes of the mixed-phase parameter $\Delta_P$, and the EoS for $\mu_< = 1000$ MeV, which would resemble the APR EoS, does not have the highest probability. 
		} 
	\end{figure}

	In Figure ~\ref{fig:BA}, we show the posterior probability distribution in the space of hybrid EoS models spanned by the values for the two model parameters 
	$\mu_<$ and $\Delta_P$ for the three observable constraints that are included in our Bayesian analysis (from left to right): 
	\begin{enumerate}
		\item[{1)}] the combined mass measurements of PSR J0740+6620 \cite{Cromartie:2019kug} and PSR J0348+0432 \cite{Antoniadis:2013pzd} as a lower limit on the maximum mass;
		\item[{2)}] this mass constraint together with the compactness constraint from tidal deformabilities of the binary compact star merger GW170817 \cite{Abbott:2018exr}; 
		\item[{3)}] additionally, the constraint from one of the mass-radius measurements by NICER \cite{Miller:2019cac}.
	\end{enumerate}
	
	\begin{figure}[H]
		\vspace{3mm}
		\includegraphics[height=0.35\textwidth]{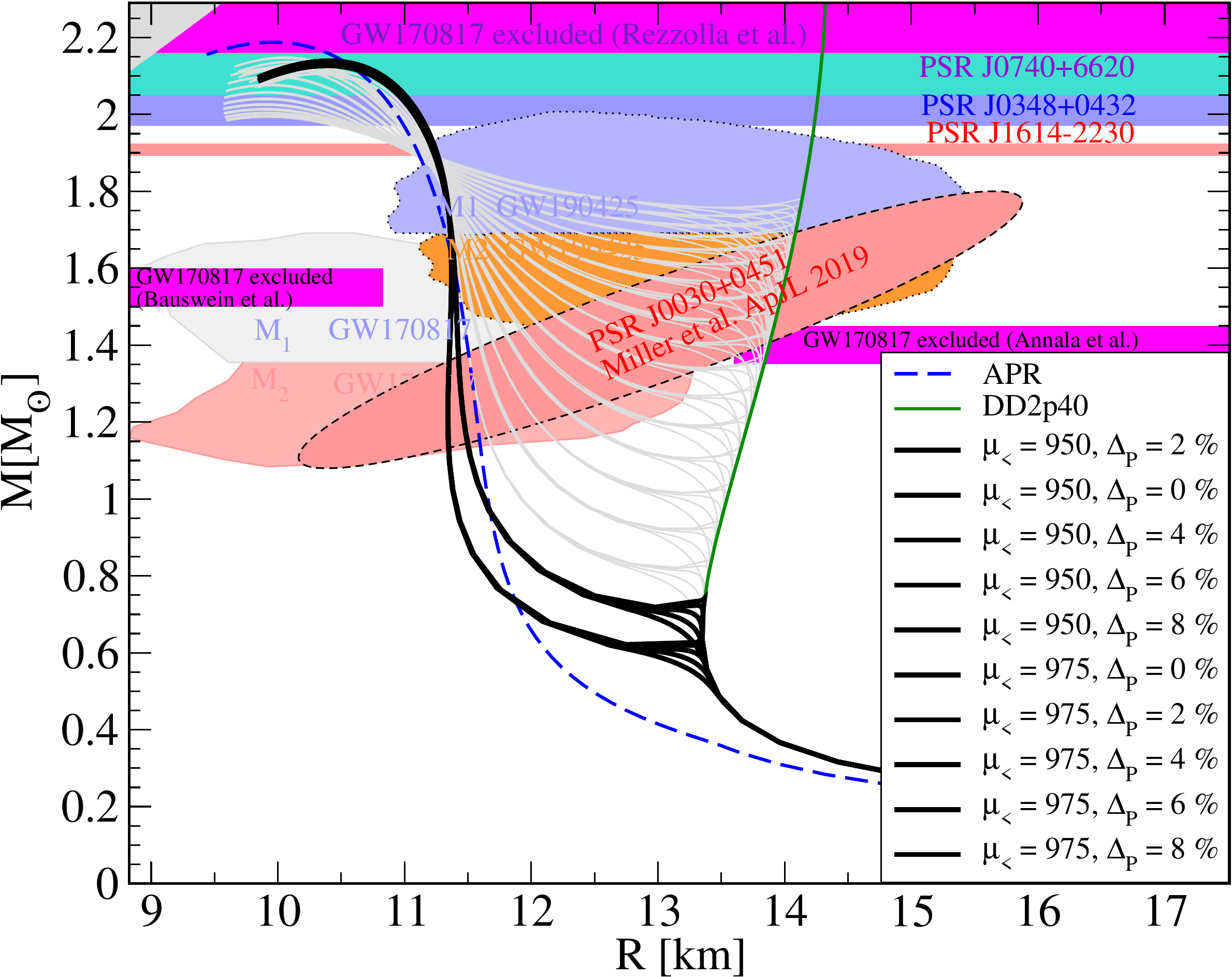} \hspace{3mm}
		\includegraphics[height=0.35\textwidth]{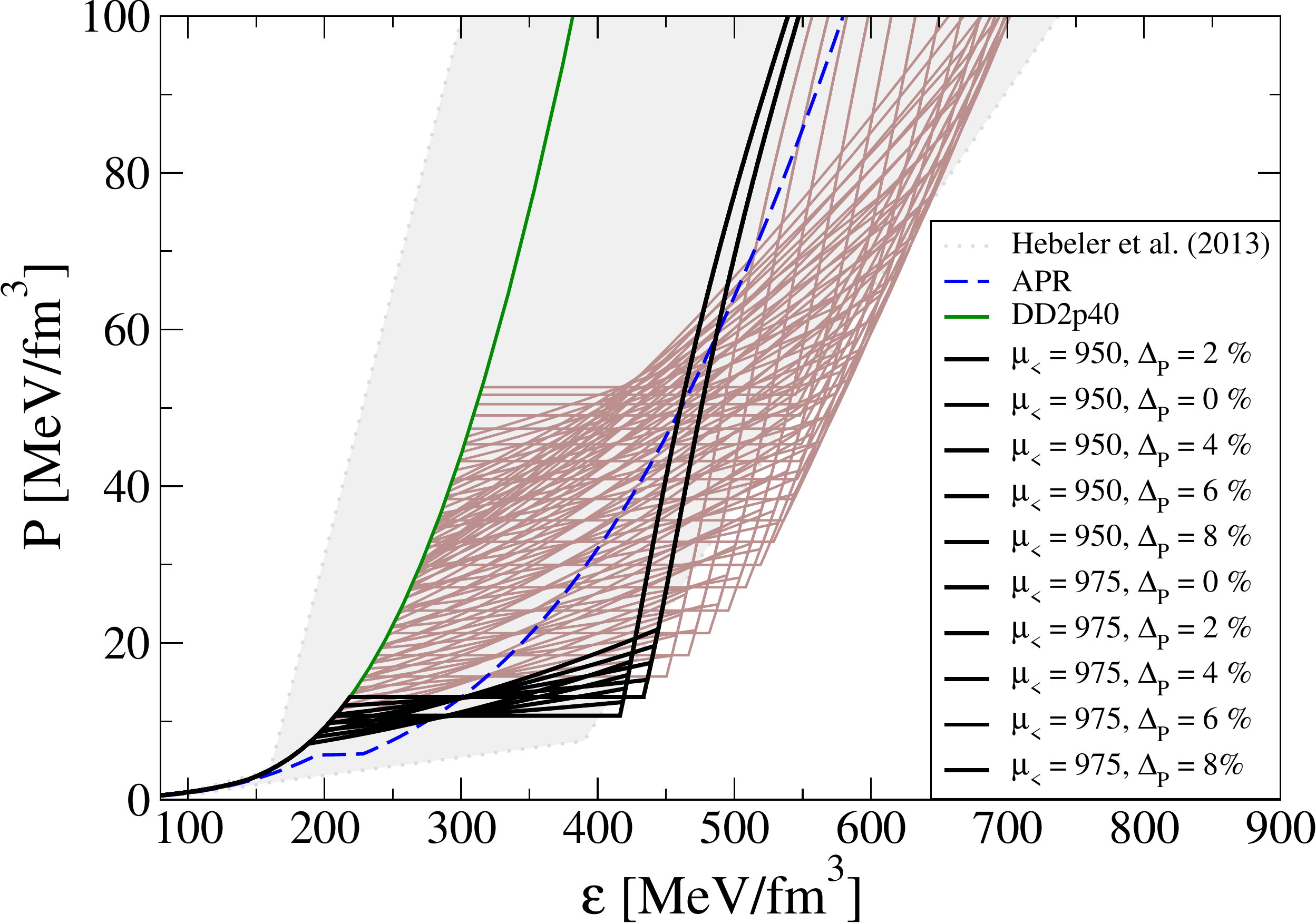} 
		\caption{\label{fig:MR-PEpost}
			Most probable compact star sequences in the $M-R$ diagram (left panel) for the most probable EoS (right panel) shown as bold black lines, corresponding to a posterior probability higher than 75 \% of the maximum one in the rightmost panel of Figure ~\ref{fig:BA}, after Bayesian analysis with the present observational constraints given by the reddish areas in Figure ~\ref{fig:NICER}; values for the parameter $\mu_<$ are in MeV.
			For orientation, we show in the right panel also the preferable region deduced from the mass-radius constraint in the pre-multi-messenger era by Hebeler et al.~\cite{Hebeler:2013nza} (grey shaded area).
		} 
	\end{figure}

	From this analysis, we deduce the most likely parameter cases, i.e., for which the posterior probability is at least 75\% of the maximal probability, and highlight them 
	in Figure ~\ref{fig:MR-PEpost} in the $M-R$ diagram (left panel) and the EoS $P(\varepsilon)$ (right panel). 
	Out of a set of 50 EoS with the same prior probability remain 10 models with a posterior probability higher than 75 \% of the maximum one, corresponding to a selectivity of 20\%, which all would favor a low onset of the deconfinement transition at $M_{\rm onset}< 0.75~M_\odot$, so that the resulting preferable EoS 
	would be almost indistinguishable from the soft hadronic APR EoS in the relevant mass range of GW170817, which is representative for the typical neutron star mass range.
	Notably, the case $\mu_< = 1000$ MeV, which would closely resemble the APR EoS, does not belong to these 10 models with the highest probability.
	
	\begin{figure}[H]
		\centering
		\includegraphics[width=0.23\textwidth]{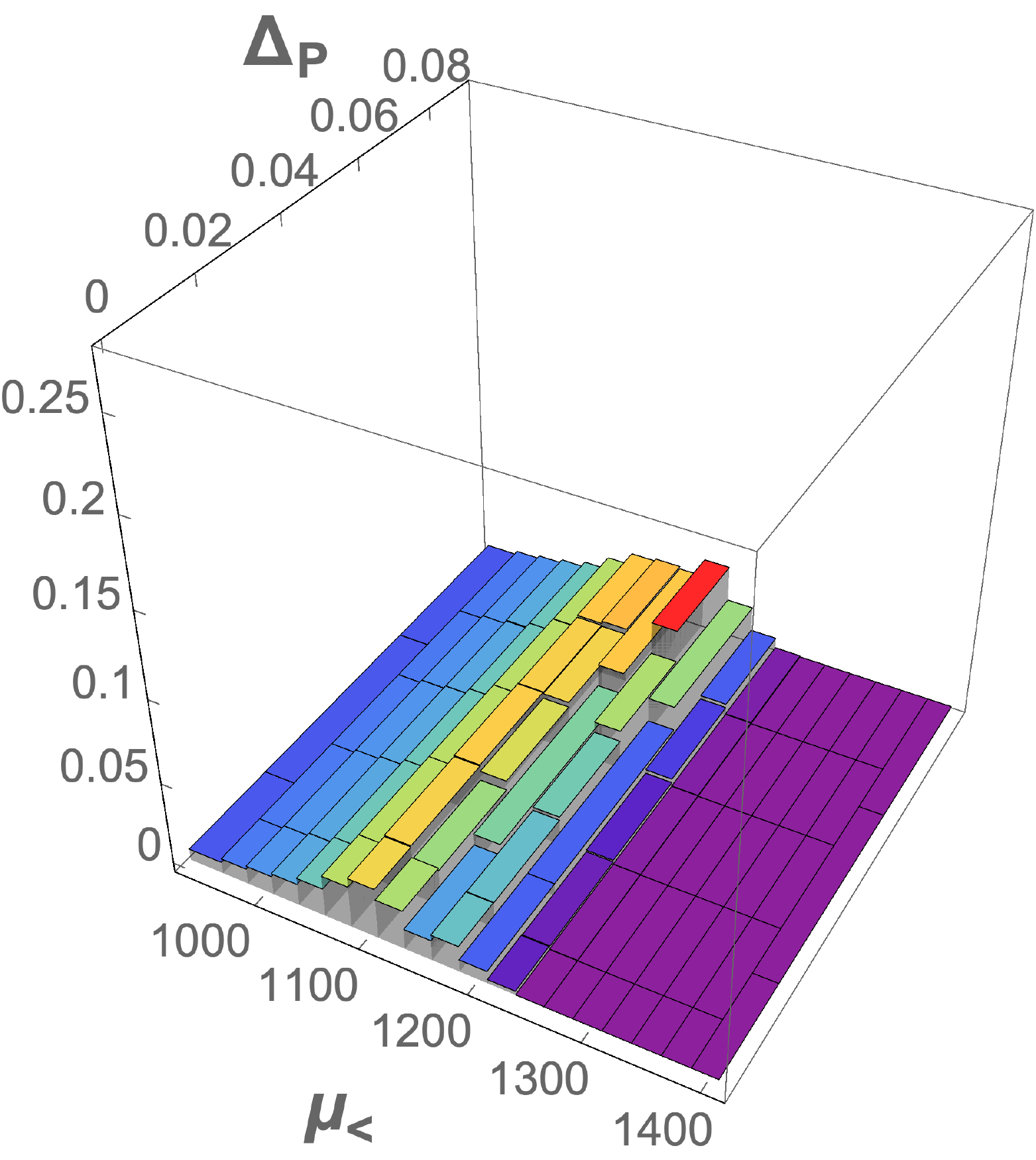} 
		\includegraphics[width=0.23\textwidth]{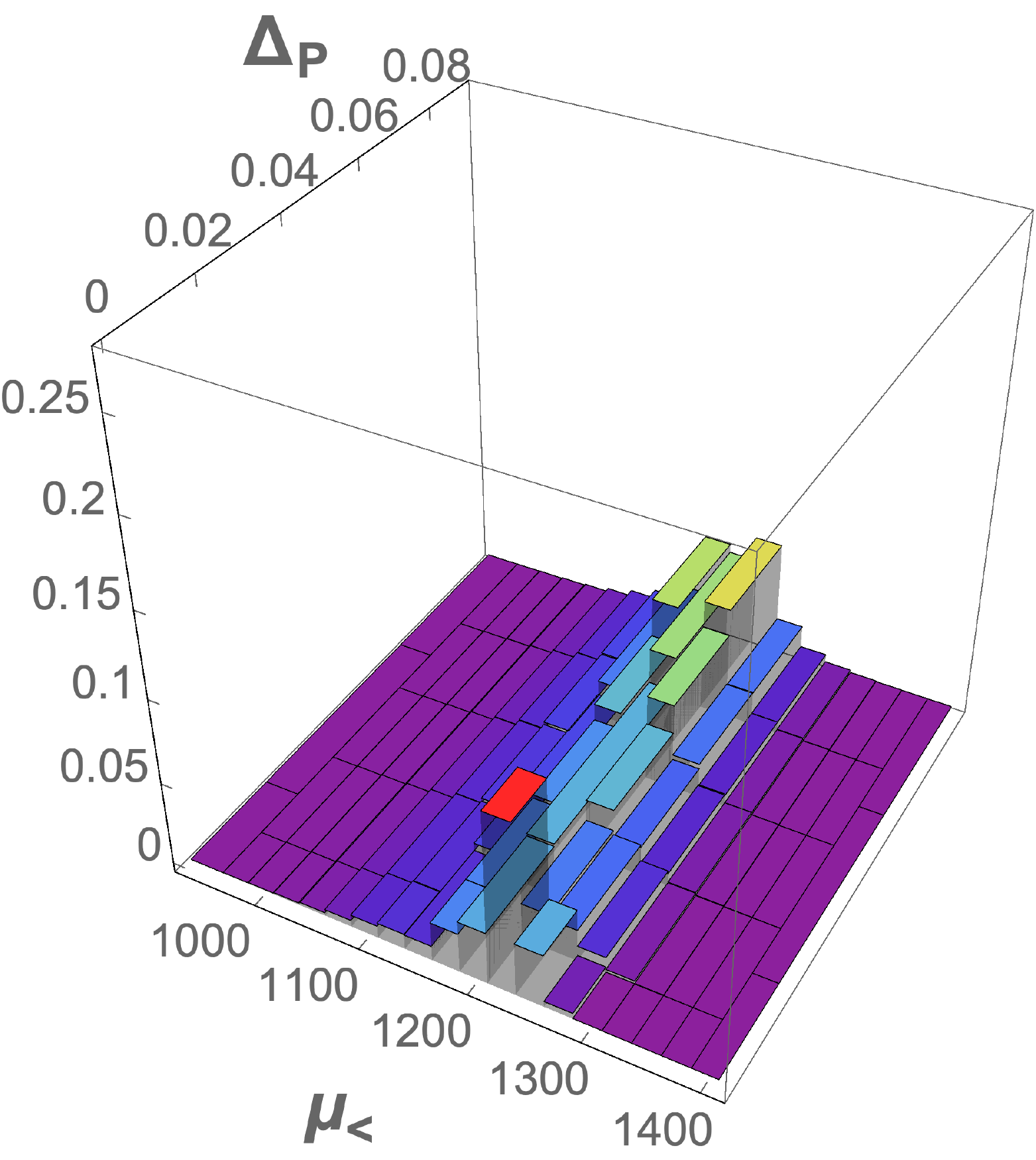} 
		\includegraphics[width=0.23\textwidth]{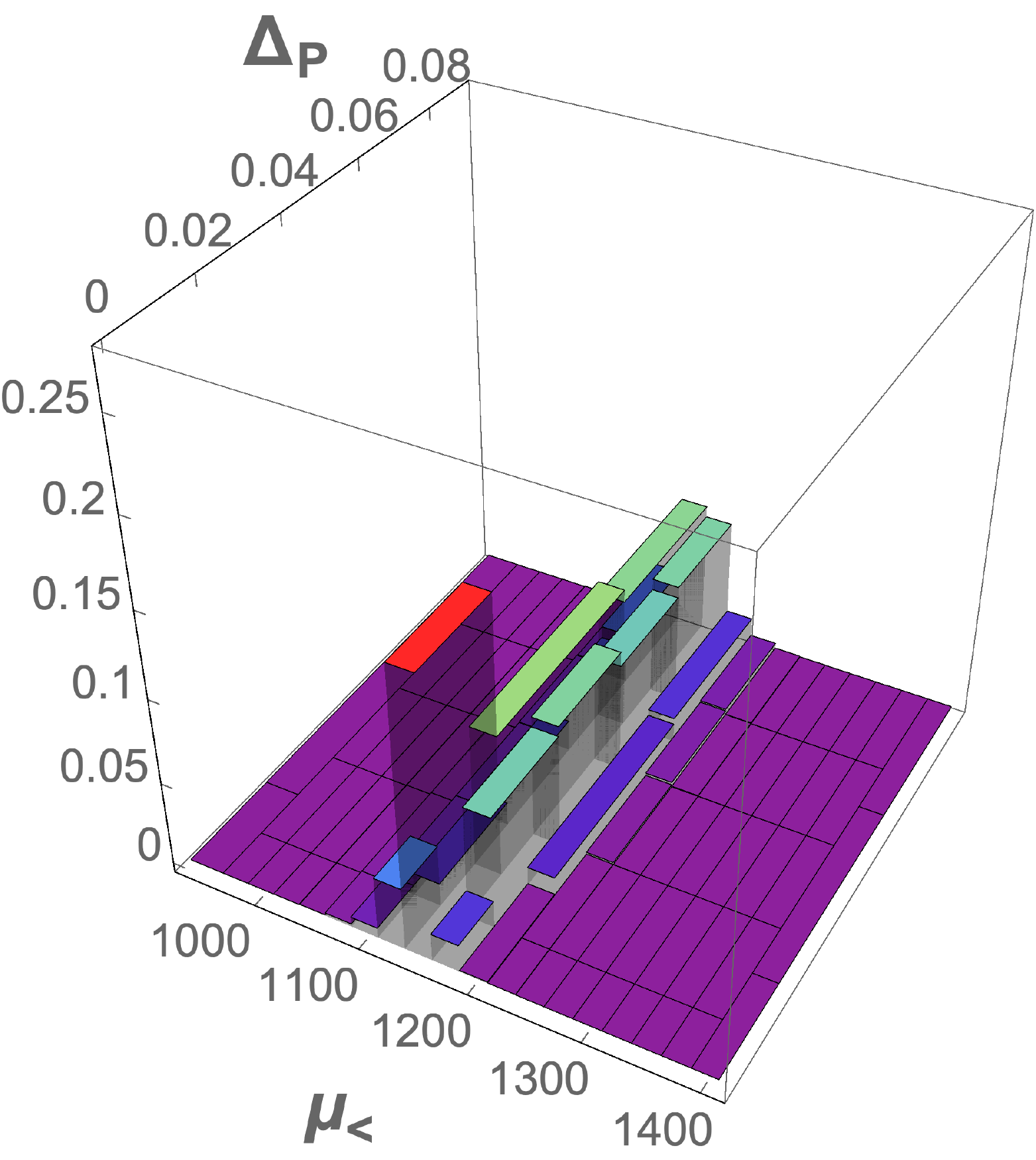} 
		\includegraphics[width=0.23\textwidth]{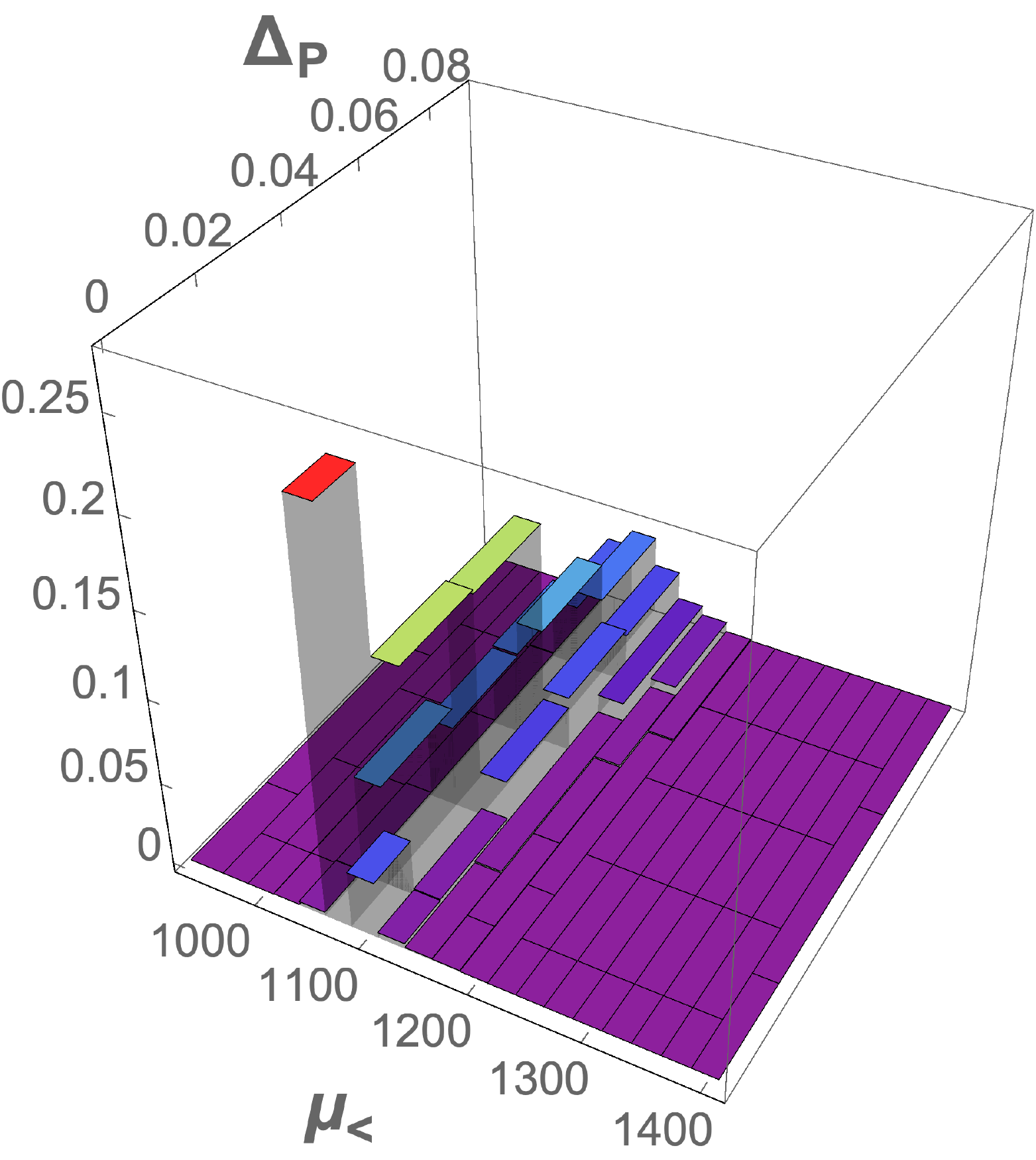} 
		\caption{\label{fig:BAfict}
			Posterior probabilities for different models after Bayesian analysis using the fictitious mass and radius measurements suggested in Figure ~\ref{fig:NICER} corresponding to possible future results from the NICER experiment, which would have a standard deviation 
			reduced by a factor two, and the mean values of mass and radius would be (from left to right): (i) unchanged, (ii) $\mu_R = 14\,\textrm{km}$, $\mu_M = 1.6~{M}_\odot$, (iii) $\mu_R = 14\,\textrm{km}$, $\mu_M = 1.4~{M}_\odot$, and (iv) $\mu_R = 14\,\textrm{km}$, $\mu_M = 1.6~{M}_\odot$.
		} 
	\end{figure}
	
\begin{figure}[H]
		\vspace{-8mm}
		\includegraphics[width=0.45\textwidth]{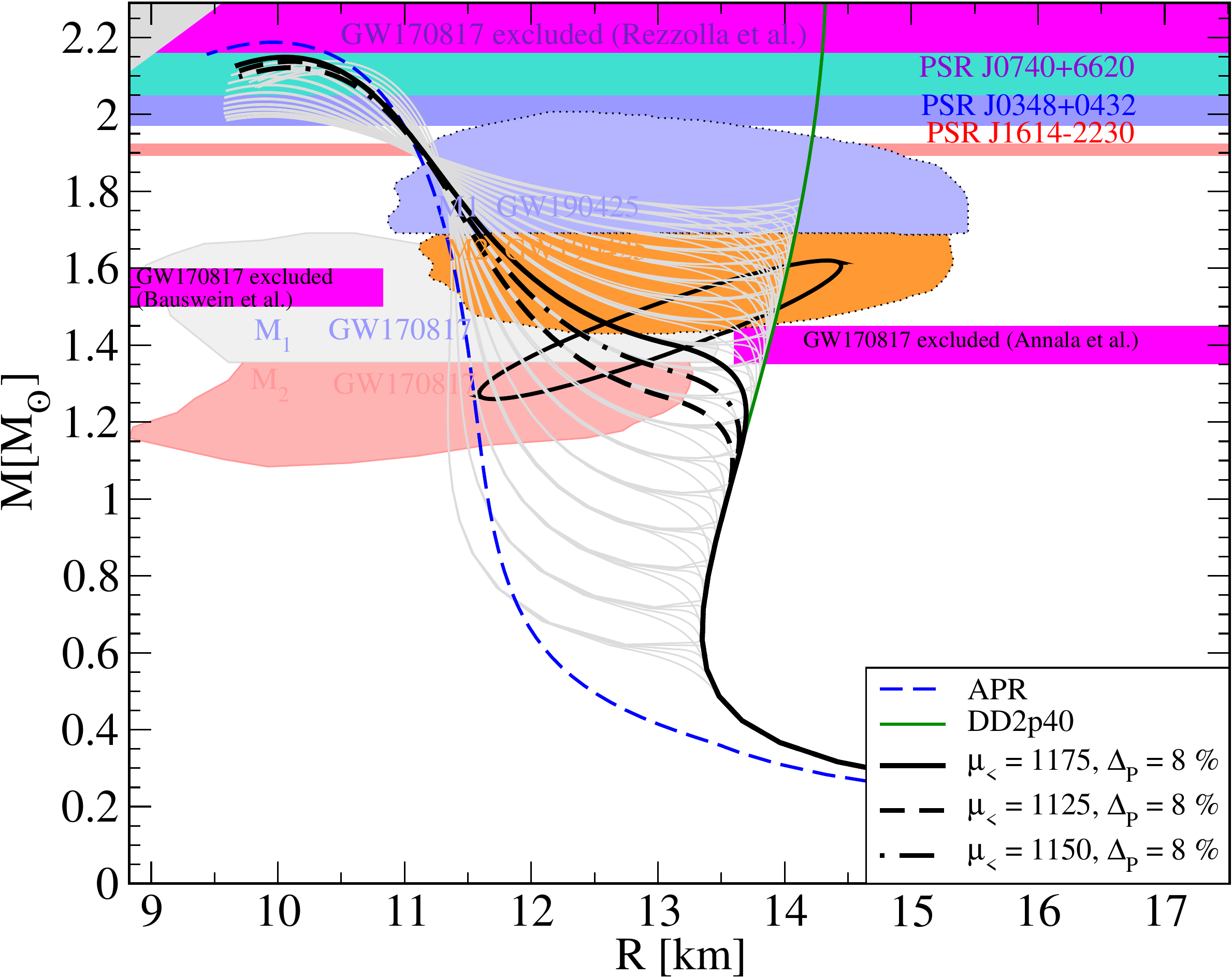} 
		\includegraphics[width=0.55\textwidth]{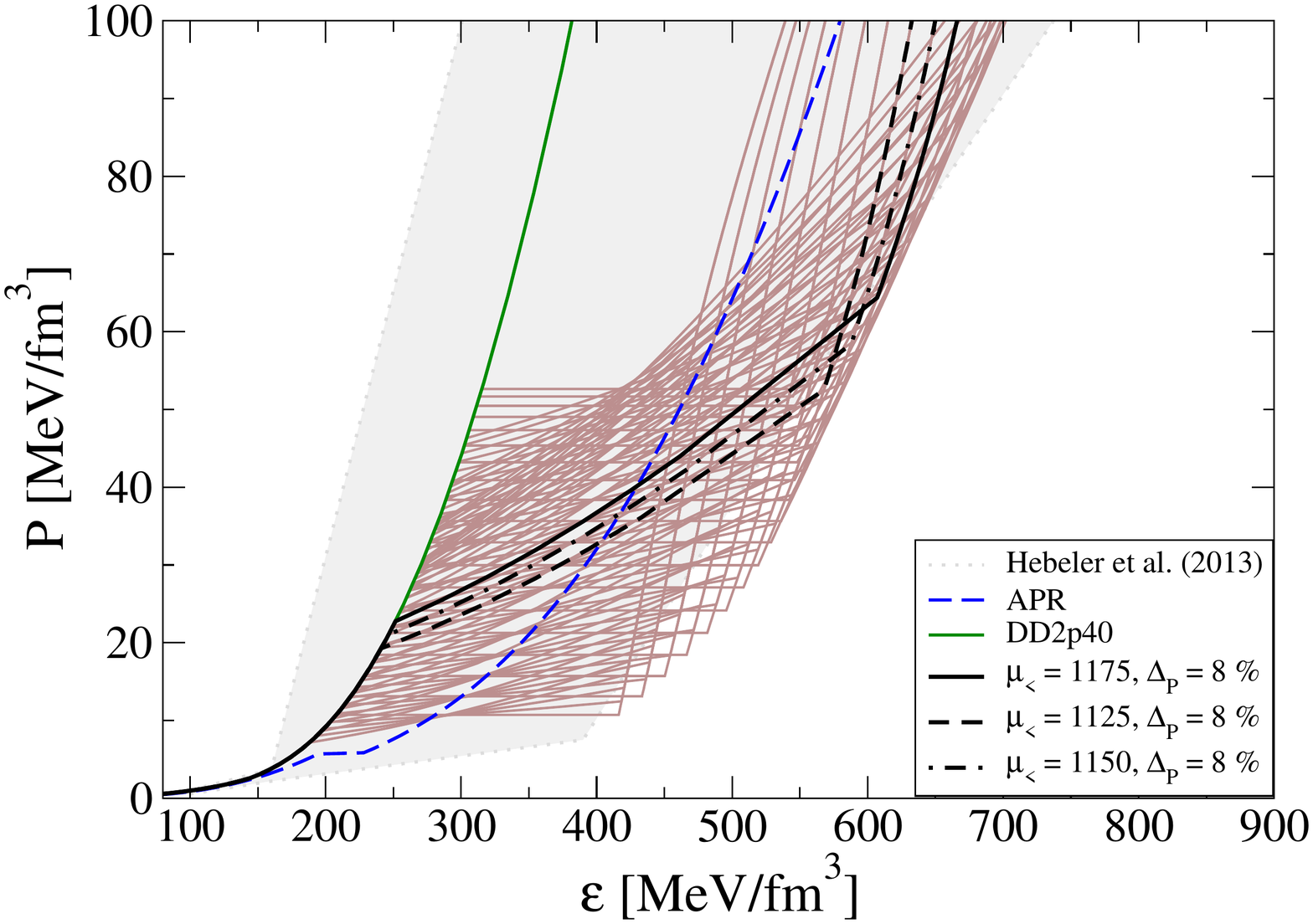} 
		\caption{\label{fig:MR-PEpost2}
			Most probable compact star sequences in the $M-R$ diagram (bold black lines in the left panel) for the most probable EoS (bold black lines in the right panel), corresponding to a posterior probability higher than 75 \% of the maximum one in the leftmost panel of Figure ~\ref{fig:BAfict}, after Bayesian analysis with a fictitious mass-radius measurement (black ellipse in the left panel) for which the standard deviation is reduced to half of the present values reported by the NICER experiment \cite{Miller:2019cac}; see Figure ~\ref{fig:NICER}; the values for the parameter $\mu_<$ are in MeV.	For orientation, we show in the right panel also the preferable region deduced from the mass-radius constraint in the pre-multi-messenger era by Hebeler et al.~\cite{Hebeler:2013nza} (grey shaded area).
		} 
	\end{figure}

	In Figure ~\ref{fig:BAfict}, we show the results of the Bayesian analysis for the four cases of fictitious mass and radius measurements, which could be provided 
	by the NICER experiment in the near future. In the leftmost panel, we demonstrate the effect that the reduction of the standard deviations of mass and radius to
	half of their present value would have. 
	Note that the planned accuracy of the radius measurement corresponded to $\sigma_M=500$ m, still smaller than what is assumed here. 
	When comparing with the rightmost panel of Figure ~\ref{fig:BA}, an increase in the selectivity, as well as in the favored onset mass of the phase transition is observed. The Bayesian analysis selects as the most likely EoS the ones with the largest mixed phase parameter, corresponding to a Gibbs construction of the phase transition with the pressure in the mixed phase ranging from 20 MeV/fm$^3$ to 60 MeV/fm$^3$.
	The favorable mass range for the onset of the deconfinement transition is $M_{\rm onset}\sim 1.1 - 1.2~M_\odot$; see Figure ~\ref{fig:MR-PEpost2}.
	
\begin{figure}[H]
		\vspace{-5mm}
		\includegraphics[width=0.55\textwidth]{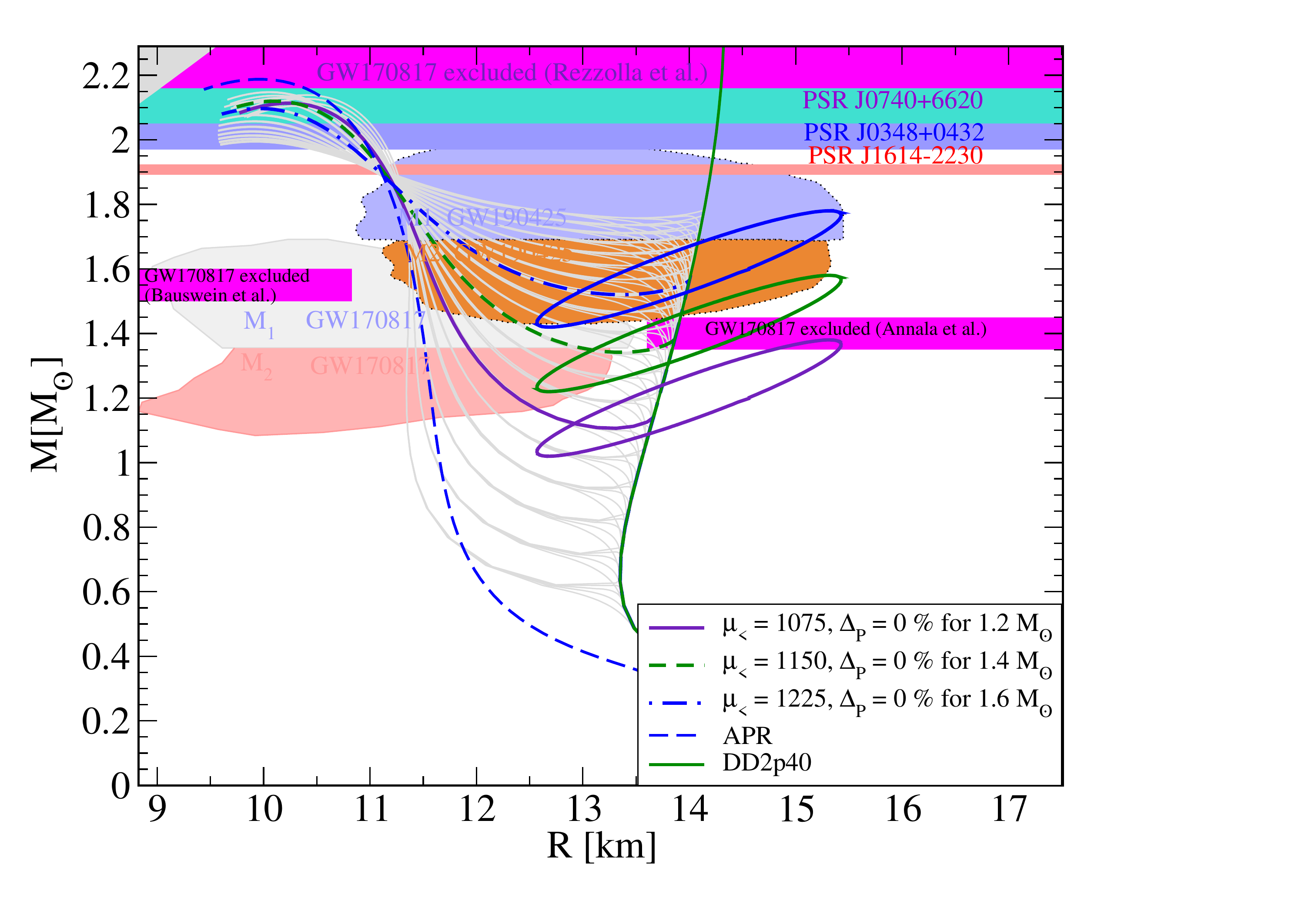} 
		\hspace{-15mm}
		\includegraphics[width=0.55\textwidth]{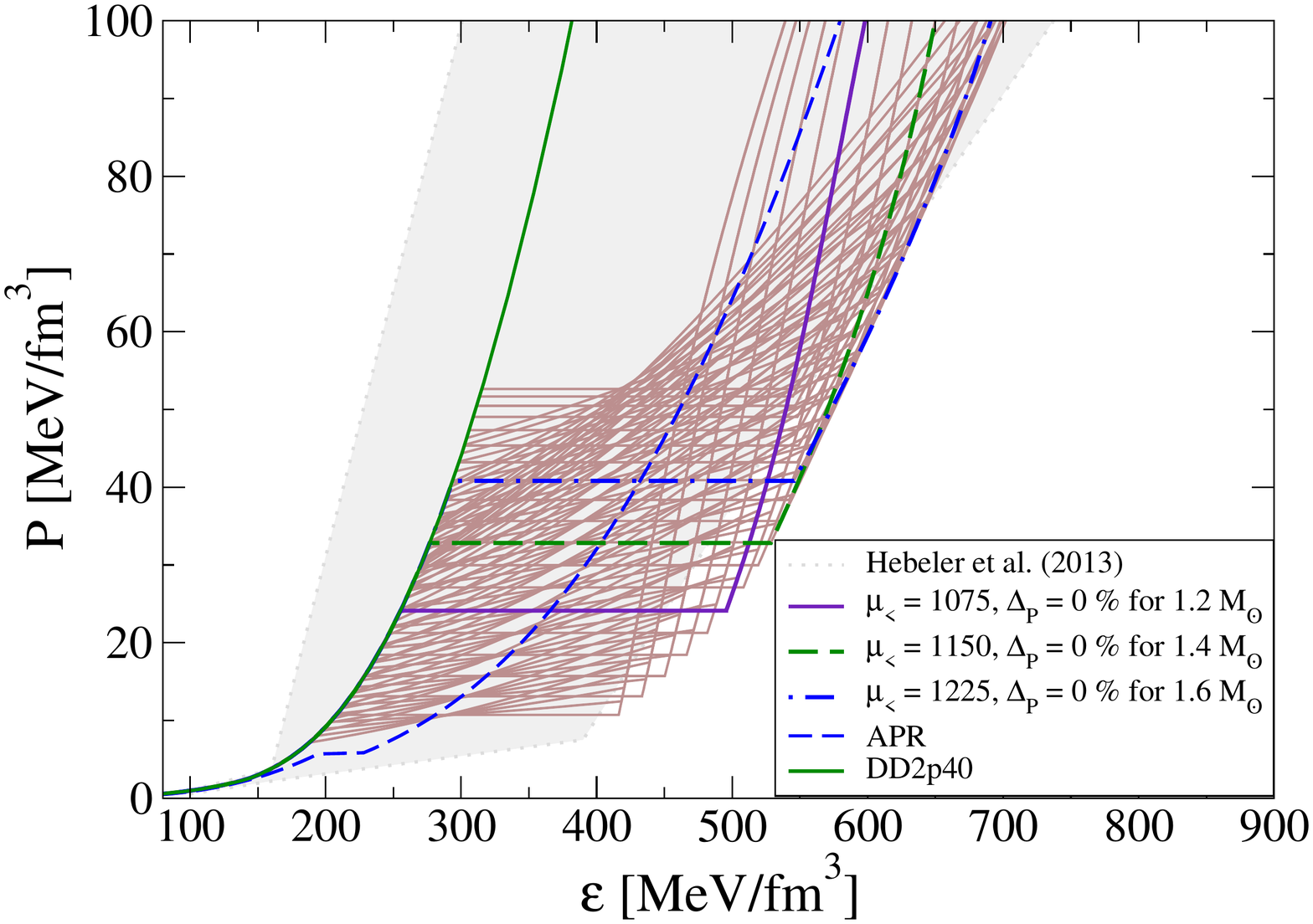} 
		\caption{\label{fig:MR-PEpost3}
			The same as Figure ~\ref{fig:MR-PEpost3}, but for a
			Bayesian analysis with fictitious mass-radius measurements (colored ellipses in the left panel) for which the standard deviation is reduced to half of the present values reported by the NICER experiment \cite{Miller:2019cac}; the radius is set to 14 km, and three masses are chosen: $1.6~M_\odot$ (blue dash-dotted line), $1.4~M_\odot$ (green dashed line), and $1.2~M_\odot$ (orange solid line); see Figure ~\ref{fig:NICER}; the values for the parameter $\mu_<$ are in MeV. For orientation, we show in the right panel also the preferable region deduced from the mass-radius constraint in the pre-multi-messenger era by Hebeler et al.~\cite{Hebeler:2013nza} (grey shaded area).		} 
	\end{figure}

	In Figure ~\ref{fig:MR-PEpost3}, we highlight the most probable compact star sequences in the $M-R$ diagram (left panel) and EoS in the pressure vs. energy density diagram (right panel) as blue, green, and orange lines corresponding to the posterior probabilities higher than 75 \% of the maximum ones in the three right panels of Figure ~\ref{fig:BAfict}, after Bayesian analysis with the fictitious mass-radius measurements (colored ellipses in the left panel) given in Figure ~\ref{fig:NICER}.
	In the case of these fictitious measurements, the Maxwell construction of the first-order phase transition with $\Delta_P=0$ is favored, and the onset mass for the deconfinement transition is strongly correlated with the mass measurement at this radius of 14.0 km, as illustrated in the left panel of Figure ~\ref{fig:MR-PEpost3}.
	
	We also explored a fictitious measurement with the narrowed error ellipse ($\sigma_M = 0.0725\, {M}_\odot$, $\sigma_R = 0.575\,\textrm{km}$) when the mass of PSR J0030+0451 was kept at $13.02~M_\odot$, but the radius was shifted to $\mu_R=12.0$ km. In this case, the Bayesian analysis exactly reproduced the 
	result of the real NICER measurement \cite{Miller:2019cac}, as given in Figure ~\ref{fig:MR-PEpost}, with the early onset of the deconfinement transition and the insensitivity to the mixed phase construction.

	\section{Conclusions}
	We developed a Bayesian analysis method for selecting the most probable equation of state under a novel set of constraints from compact star observations, 
	which now include, besides the actual lower limit on the maximum mass from PSR J0740+6620 and the tidal deformability from GW170817, also the first result for the mass and radius
	reported by the NICER Collaboration, for PSR J0030+0451.
	We applied this method for the first time to a two-parameter family of hybrid equations of state that was based on realistic models for hadronic matter (DD2 with nucleonic excluded volume) and color superconducting quark matter (generalized nlNJL model), which produce a third family of hybrid stars in the mass-radius diagram.
	
	It turned out that an early phase transition was favored, and the results were rather independent of the mixed phase parameter, i.e., whether the hybrid stars formed a third family branch or the hybrid star sequence was connected with the hadronic sequence for sufficiently large $\Delta_P>4~\%$. 
	The presence of multiple configurations for a given mass (twins or even triples) corresponded to a set of disconnected lines in the $\Lambda_1-\Lambda_2$ diagram of tidal deformabilities for binary mergers, so that merger events from the same (sufficiently narrow) mass range may result in probability landscapes in this diagram with different peak positions corresponding to different origins of the stars in the binary. 
	
	The hybrid equation of state favored under this set of present observables corresponded to a line in the $M-R$ diagram, which was very similar to that for the APR equation of state, which it practically resembled in the range of observed neutron star masses ($1.1 \dots 2.1~M_\odot$). 
	This was a modern reappearance of the masquerade problem discussed first in \cite{Alford:2004pf}.
	The situation would change dramatically if NICER would improve the accuracy of its mass and radius determination for J0030+0451 so that it would have only half of the present variance.
	In this case, solutions with the onset mass for deconfinement in the range between 1.1 and 1.2 $M_\odot$ would be favored that have a large mixed phase parameter and do not exhibit mass twins. 
	If additionally, the radius would be as large as 14 km, then there remains just one favorable solution with a Maxwell construction and mass twins for 
	an onset mass that is correlated with the measured mass value. We showed the examples for 1.2, 1.4, and 1.6 $M_\odot$. 
	If with this reduced variance the radius would be measured to 12 km, then the favorable solutions would be the same as for the actual NICER measurement, exhibiting an early onset of the deconfinement transition at masses below $0.8~M_\odot$, insensitive to the mixed phase parameter $\Delta_P$.
	Our ongoing work explores other phase transition constructions, such as a crossover type interpolation between a soft hadronic EoS and a stiff high-density quark matter EoS as the original nlNJL model. 
	Such a construction allows discussing softer hadronic EoS for the hybrid EoS construction as, e.g., the APR or DD2F equations of state.
	We are also investigating the hyperon puzzle \cite{Logoteta:2019utx,Lonardoni:2014bwa}
	in this context \cite{Shahrbaf:2019vtf,Shahrbaf:2020uau}.
	Finally, there are systematic studies ongoing to unify the hadronic and quark matter descriptions 
	\cite{Bastian:2018wfl,Marczenko:2018jui,Bastian:2018mmc,Marczenko:2020jma}. 
	A review-type publication of our results is in preparation. 
	
	
	\vspace{6pt} 
	
	\authorcontributions{{Conceptualization and methodology, A.A., D.B., and H.G.; software, A.A. and D.E.A.-C.; 
			investigation, A.A.; 
			data curation, A.A. and D.E.A.-C.; 
			writing, original draft preparation, A.A., D.B., and D.E.A.-C.; writing, review and editing, D.B. and H.G.; visualization, A.A.; supervision, H.G.; project administration, D.B.; funding acquisition, D.B. and D.E.A.-C. All authors have read and agreed to the published version of the manuscript}} 
	
	
	\funding{{A.A., D.B., and H.G. acknowledge support from the Russian Science Foundation under Grant No. 17-12-01427 for the work described in Sections 2.2 and 3-5. The work of D.B. in Subsections 2.1 and 2.3 was supported by the Polish National Science Centre under Grant Number UMO 2019/33/B/ST9/03059.
			D.E.A.-C. is grateful for partial support from the Ter-Antonian--Smorodinsky program for collaboration between JINR Dubna
			and Armenian scientific institutions and from the Bogoliubov--Infeld program for collaboration between JINR Dubna and Polish institutions.}
	}
	
	\acknowledgments{We acknowledge Stefan Typel for providing the equation of state data of the DD2p40 model and Gabriela Grunfeld for those of the nlNJL 
		model in tabulated form. 
		We thank Cole Miller for his comments on a draft of this paper, {Konstantin} 
		Maslov for discussions on the hybrid star EoS, and Sergei Blinnikov for his remark on the special point in the M-R diagram with Ref.~\cite{Yudin:2014mla}.
		The authors acknowledge the COST Actions CA15213 "THOR" and CA16214 "PHAROS" for supporting their networking activities.
	}
	\conflictsofinterest{The authors declare no conflict of interest.}
	
	
	\reftitle{{References}}
	

\end{document}